\documentclass[10pt,english,oneside]{article}

\usepackage{fancyhdr}
\usepackage{amssymb,latexsym}
\usepackage{amsmath,amsthm}
\usepackage{amsfonts}
\usepackage[latin1]{inputenc}
\usepackage{cite}

\newcommand{\sss}{\setcounter{equation}{0}}

\fancypagestyle{plain}{\fancyhead{}}

\newtheorem{theorem}{THEOREM}[section]

\newtheorem{lemma}[theorem]{LEMMA}

\newtheorem{remark}[theorem]{REMARK}

\newtheorem{prop}[theorem]{PROPOSITION}

\newtheorem{definition}[theorem]{DEFINITION}

\def\R{\mathbb{R}}
\def\Rn{\mathbb{R}^n}
\def\Hscr{{\cal H}}
\def\Lscr{{\cal L}}
\def\Vscr{{\cal V}}

\def\Sscr{{\cal S}}

\def\scero{_0}
\def\suno{_1}
\def\sdos{_2}

\def\euno{^1}
\def\edos{^2}

\def\bX{\hbox{{\bf X}}}
\def\hbX{{\hat{\bf X}}}
\def\bP{\hbox{{\bf P}}}

\def\tbp{{\tilde{\bf p}}}
\def\tbx{{\tilde{\bf x}}}
\def\tbR{{\tilde{\bf R}}}

\def\bd{\hbox{{\bf d}}}
\def\be{\hbox{{\bf e}}}
\def\bE{\hbox{{\bf E}}}
\def\bp{\hbox{{\bf p}}}
\def\bq{\hbox{{\bf q}}}
\def\bv{\hbox{{\bf v}}}
\def\bu{\hbox{{\bf u}}}
\def\bx{\hbox{{\bf x}}}
\def\by{\hbox{{\bf y}}}

\def\bsv {{\scriptstyle {\bf v} }}
\def\bsx{{\scriptstyle {\bf x} }}
\def\bsy{{\scriptstyle {\bf y} }}
\def\bsp{{\scriptstyle {\bf p} }}

\def\bse{{\scriptstyle {\bf e} }}
\def\bsE{{\scriptstyle {\bf E} }}

\def\(({\hbox{\bf (}}
\def\)){\hbox{\bf )}}

\def\sj{_{j}}
\def\sk{_{k}}
\def\jp{ j^{\prime}}
\def\kp{ k^{\prime}}
\def\sjk{ _{jk }}
\def\sjpkp{ _{j^{\prime} k^{\prime}  }}

\def\beq{\begin{equation}}
\def\ene{\end{equation}}

\def\bull{\vrule height 6pt width 6pt depth -.pt} 

\def\fin{\hbox{\vrule height6pt width6pt depth0pt}}

\newenvironment{dem}{\textsc{Proof.\quad}}{ \hfill \fin}

\begin{document}

\title{High-Velocity Estimates and Inverse Scattering for Quantum N-Body Systems with
Stark Effect  \thanks{ PACS Classification (2010): 03.65.Nk, 03.65.Ca, 03.65.Db, 03.65.Ta.  Mathematics Subject Classification(2010): 81U40, 35P25,
35Q40, 35R30.}}
\author{ Gerardo Daniel Valencia\\
Departamento de Física Matemática.\\
Instituto de Investigaciones en
Matem\'aticas Aplicadas y en Sistemas.\\
Universidad Nacional
Aut\'onoma de M\'exico.\\
Apartado Postal 20--726. M\'exico, D.F.
01000.\\
e-mail:
gvalenciamx@yahoo.com.mx\\
\and Ricardo Weder \thanks { Fellow Sistema
Nacional de Investigadores. Research partially supported by CONACYT under grant CB2008-I 99100.} \thanks {On leave of absence Departamento de Física Matemática.
Instituto de Investigaciones en
Matem\'aticas Aplicadas y en Sistemas.
Universidad Nacional
Aut\'onoma de M\'exico.
Apartado Postal 20--726. M\'exico, D.F.
01000.} \\
Projet POems, Domaine de Voluceau-Rocquencourt, Institut National de Recherche en \\
Informatique et en Automatique Paris-Rocquencourt, BP 105, 78153 Le Chesnay Cedex \\
France. \\
e-mail: weder@unam.mx}

\date{\today}

\maketitle

\begin{center}
\begin{minipage}{5.75in}

\centerline{{\bf Abstract}} \bigskip

In an $N$--body quantum system with a constant electric field, by inverse scattering, we uniquely reconstruct pair potentials, belonging to the optimal class of short-range potentials and long-range potentials, from the high-velocity limit of the Dollard scattering operator. We give a reconstruction formula with an error term.

\end{minipage}
\end{center}

\section{Introduction}\sss

We study the direct and inverse scattering problems for an
$N$--body quantum mechanical system in an $n \geq 2$ dimensional
space
under Stark effect, i.e. in a constant electric field,
with interactions given by pair
potentials (multiplication operators).

When we speak of scattering by a potential $V$, it is common that $V$ is classified as being short-range if the canonical wave operators $W_\pm(H\scero+V,H\scero)$ exist, where $H\scero$ is the unperturbed Hamiltonian. On the other hand, if they do not exist, we say that we have a long-range potential; in this case we have to modify the free evolution and thus, to define modified wave operators.

As it is well known, the Coulomb potential $V_c (x) = q/|x|$ is long-range when $H\scero = -\triangle$. It is also well known, that $V_c$ is short-range in the case of the Stark effect, where $H\scero = -\triangle - E \cdot x,$ and $E$ is a constant electric field. More generally, potentials $V$ that decay at infinity as $V(x) \approx |x|^{-\gamma},$ $\gamma \leq 1$ are long-range when $H\scero=-\triangle$ and on the contrary, when there is a constant electric field, they are short-range if $1/2 < \gamma \leq 1$ and long-range if $0 < \gamma \leq 1/2.$

This feature of the Stark effect is particularly interesting in inverse scattering. For example, because it allows to prove that the Coulomb potential is uniquely determined by the scattering matrix, defined from canonical wave operators, without having to modify the free dynamics, as first proved in \cite{Weder1996MISCIAEF}.

We denote by $m_j \in \R^+, q_j \in \R$  and $\tbx_j \in \R^n, j = 1,2, \ldots, N$,
respectively, the masses, the charges and the positions of the particles. The free
Hamiltonian generates the free time evolution,
\beq\label{Art1eq 1.9}
\tilde H\scero = \sum\limits^N_{j=1} (2 m_j)^{-1} \ \tbp^2_j \ + \sum\limits^N_{j=1} q\sj \bE \cdot \tbx\sj,
\quad
\tbp_j = -i \nabla_{\tbx_j},
\ene
where the electric field $\bE = (-E, 0, \ldots, 0)$, $E = |\bE|>0$ is directed along minus the first coordinate direction.

We study the system in the center of mass frame and we separate off the motion of the center of mass
$$H_{C M} = \left( 2 M \right)^{-1}
\left( \bP_{CM}\right)^2 + Q\bE \cdot \bX_{CM},$$
where $M=\sum\limits^N_{j = 1} m\sj$, is the total mass, $
\bX_{CM}=(1/M)\sum\limits^N_{j = 1} m\sj \tbx\sj$, is the center of mass,
$\bP_{CM}=\sum\limits^N_{j = 1} \tbp\sj$, is the momentum of the center of mass, $Q=\sum\limits^N_{j = 1} q\sj,$ is the total charge.

The free Hamiltonian in the center of mass frame is $H \scero: = \tilde H\scero - H_{CM}$,

$$H\scero = \sum\limits^N_{j=1} (2 m_j)^{-1} \ \tbp^2_j \ - \left( 2 M \right)^{-1} \left( \bP_{CM}\right)^2 + \sum\limits^N_{j=1} (q\sj-m\sj Q/M) \bE \cdot \tbx\sj.$$

$H\scero$ is essentially self-adjoint in the space of Schwartz. We also denote by $H\scero$ the unique self-adjoint extension.
%
%

In the center of mass frame the space of states is the
Hilbert space, ${\cal \Hscr}$, represented in configuration space by
wave functions $\phi$ in
\beq\label{Art1eq LX}
L\edos (\bX),
\quad
\bX = \left\{ \tbx=(\tbx\suno, \ldots, \tbx_N) \bigg| \sum^N_{j=1}
m_j\, \tbx_j = 0 \right\} \cong \R^{n(N-1)}
\ene with the measure induced on \bX \ by the following norm on  $\R^{nN} :
|||\tbx|||= \left[ \sum\limits^N_{j=1} m_j\, \tbx_j\edos\right]^{1/2}$.
The space
\beq\label{Art1eq LX_hat} L\edos (\hat{\bf X}),\quad \hat{\bf X} = \left\{\tbp=(\tbp\suno, \ldots,
\tbp_N) \bigg| \sum^N_{j=1} \tbp_j = 0\right\} \cong \R^{n (N-1)},
\ene where $\hbX$ is equipped with the dual metric induced by
$\left[\sum\limits^N_{j=1} (m_j)^{-1}\,\tilde{\bp}\edos_j\right]^{1/2}$ on
$\R^{nN},$ is the set of momentum space wave functions $\hat \phi$.
Fourier transform maps unitarily $L\edos (\bX)$ onto $L\edos (\hbX)$.
The measures on $\bX$ and $\hbX$ are equivalent to Lebesgue measure.
Given an (abstract) state $\Phi \in \Hscr$ we use both its
configuration or momentum space wave functions where appropriate.

As a general reference for multiparticle scattering see e.g.\ \cite{ReedSimonIII}, where Jacobi coordinates are defined

\beq\label{Art1eq 1.12}
\xi\sj := \tbx_{j+1} - \left( \sum_{k = 1}^j m\sk \right)^{-1} \left( \sum_{k = 1}^j m_k \tbx\sk \right), \quad j=1, \ldots, N-1.
\ene

These coordinates are obtained by first changing variables from $(\tbx_1, \tbx_2)$ to $\xi_1=\tbx_2 - \tbx_1$ and the center of mass of particles (1) and (2), $\tbR_{12}=(m_1+m_2)^{-1} (m_1 \tbx_1 + m_2 \tbx_2)$. Then we change from $(\tbR_{12} , \tbx_3)$ to $\xi_2=\tbx_3 - \tbR_{12}$ and the center of mass of particles (1), (2), and (3), and so on. In the end we obtain the Jacobi coordinates $\xi\sj, 1\leq j \leq N-1$, on $\bX$ and the center of mass coordinate $\bX_{CM}$. In these coordinates $H\scero$ is expressed as

\beq\label{Art1eq 1.13}
H\scero = \sum_{j=1}^{N-1} ((2 \nu\sj)^{-1} \hat{\bp}\sj^2 + q\sj^R \bE \cdot \xi\sj), \quad \hat{\bp}\sj = -i \nabla_{\xi\sj},
\ene
where
\[
\nu\sj^{-1} = m_{j+1}^{-1} + \left( \sum_{k=1}^j m\sk \right)^{-1}, \quad 1 \leq j \leq N -1,
\]
\beq\label{Art1eq 1.14}
q\sj^R = (q_{j+1} M\sj - m_{j+1}Q_j)/(m_{j+1}+M\sj), \quad M\sj = \sum_{k=1}^j m\sk,
\ene
\[
Q\sj = \sum_{k=1}^j q\sk, \quad 1 \leq j \leq N -1,
\]
$\nu\sj$ and $q\sj^R, 1 \leq j \leq N -1$, are, respectively, the reduced mass and the relative charge of the particle $(j+1)$ with respect to the masses and the charges of the first j particles. Formula \eqref{Art1eq 1.13} shows that the proof that $H\scero$ is essentially self-adjoint in the space of Schwartz reduces to the one in the two-body case. The Jacobi coordinates above are based in the pair of particles (1, 2) in the sense that we have taken as the first coordinate $\xi_1=\tbx_2 - \tbx_1$ the relative distance of the particles (1) and (2). Of course, we can base Jacobi coordinates in any pair of particles $(j, k), j, k=1, 2, ..., N.$

In order to determine the potential for a given pair we number the particles in such a way that the given pair consists of particles one and two. By \eqref{Art1eq 1.13} we write
\beq\label{Art1eq 1.15}
H\scero = \left[ (2\nu_1)^{-1} \hat{\bp}_1^2 + \frac{(q_2 m_1 -m_2 q_1)}{m_1 + m_2} \bE \cdot \xi_1 \right] \otimes I + I \otimes \hat {H}\scero,
\ene
under the decomposition of $L^2(\bX)$ as
\[
L^2(\bX) = L^2(\R_{\xi_1}^n) \otimes \left[ \otimes \prod_{j=2}^{N-1} L^2(\R_{\xi\sj}^n)  \right],
\]
where
\beq\label{Art1eq 1.16}
\hat{H}\scero = \sum_{j=2}^{N-1} ((2 v_j)^{-1} \hat{\bp}\sj^2 + q\sj^R \bE \cdot \xi\sj).
\ene

This shows that if the relative charge of the pair (1, 2), $(q_2m_1-m_2q_1)/(m_1+m_2)$, is different from zero the relative distance of the pair (1, 2) is accelerated by the electric field as in the two-body case. However, if the relative charge is zero both particles are accelerated in the same way by the electric field and the relative distance is not accelerated, and then, with respect to the pair (1, 2), the relative scattering is as in the case when the external constant electric field is zero. This shows that, for any given pair of particles, the inverse scattering problem has to be formulated as in the two-body case with no electric field if the relative charge of the pair is zero and, as in the two-body case with an electric field, if the relative charge of the pair is different from zero.

For any given pair of particles we construct as in Enss and Weder \cite{EnssWeder1995} appropriate states where all particles have high-velocity relative to each other in order to reconstruct the corresponding pair potential. For this purpose we first introduce some kinematical notation. We use a numbering of the particles such that the pair of interest consists of particles 1 and 2. As usual we take as one n-dimensional variable the relative distance $\bx$ and momentum $\bp$ of the chosen pair $(1,2)$.

\beq\label{Art1eq 3.1}
\bx = \xi_1 := \tbx_2 - \tbx_1, \qquad \bp = \hat\bp_1 =-i \nabla_{\bsx} = \mu_{12} [ (-i \nabla_{\tbx_2}/m_2) - (-i \nabla_{\tbx_1})/ m_1) ],
\ene
where $\mu_{12}$ is the reduced mass of the pair $(1,2)$, $\mu_{12} = m_1 m_2 / (m_1 + m_2)$. We also use the position $\bx_j$ and the momentum $\bp_j$ of the jth particle, $j=1, ..., N$, relative to the center of mass of the pair (1,2),

\begin{eqnarray}
\bx_j &:=& \tbx_j - (m_1 \tbx_1 + m_2 \tbx_2) / (m_1 + m_2), \quad j=1, \ldots, N, \label{Art0eq 4.1}
\\
\bp_j &=&  \mu_j (\tbp_j / m_j - (\tbp_1 + \tbp_2) / (m_1 + m_2)), \quad j=1, \ldots, N, \label{Art0eq 4.2}
\end{eqnarray}
where $\mu_j$ is the reduced mass of the jth particle with respect to the center of mass of the pair (1,2),
\[
\mu_j = m_j (m_1 + m_2) / (m_j + m_1 + m_2), \quad j=1, \ldots, N,
\]
and $\tbp_j = -i \nabla_{\tbx_j}$ is the momentum relative to some origin (see \eqref{Art1eq 1.9}). Note that $\bx$ is the first Jacobi coordinate $\xi_1$, $\bp = \mu_{12} ( \tbp_2/m_2 - \tbp_1/ m_1 )$ and $\bp_j / \mu_j $ are the relative velocities with respect the center of mass of the distinguished pair $(1,2).$ $\{ \bx, \bx_3, \ldots \bx_N \}$ and $\{ \bp, \bp_3, \ldots \bp_N \}$ are sets of $N-1$ independent n-dimensional variables in the configuration and momentum space, respectively, relative to the center of mass frame.

Let $\Phi_0 \in \Hscr$ be an asymptotic configuration with the product wave function of the form in momentum space,
\beq\label{Art1eq 3.2}
\Phi_0 \sim \hat{\phi}_{12}(\bp) \hat{\phi}_{3}(\bp_3, \ldots , \bp_N),
\ene
where $\hat{\phi}_{12} \in C_0^\infty (\R^n)$ varies while $\hat{\phi}_{3} \in C_0^\infty (\R^{n(N-2)})$ is a fixed normalized function with support in $\{ (\bp_3, \ldots , \bp_N): \vert \bp_j \vert < \mu_j \}$; i.e., the particles 3 to N have speed smaller than one relative to the pair (1,2). We take an $\eta>0$ such that $\hat{\phi}_{12} \in C_0^\infty(B_{\mu_{12} \eta }),$ where $B_{\mu_{12} \eta}$ denotes the open ball of center zero and radius $\mu_{12} \eta$ in $\Rn$.

The high-velocity state is defined as (see Enss and Weder \cite{EnssWeder1995})
\beq\label{Art1eq 3.3}
\Phi_{\bsv} \sim \hat{\phi}_{12}(\bp-\mu_{12}\bv) \hat{\phi}_{3}(\bp_3 - \mu_{3}\bv_3, \ldots , \bp_N - \mu_{N}\bv_N),
\ene
where $\bv=v \hat{\bv}, \vert \hat{\bv} \vert = 1, \bv_j = v^2 \bd_j$, with $\bd_j \neq 0$, for $j=3, \ldots, N$ and where we assume that $\bd_j - \bd_k \neq 0$ for $j,k = 3, \ldots N$. We, moreover, define $\bv_1 = -\bv \mu_{12}/m_1, \bv_2 = \bv \mu_{12}/m_2.$

We denote the relative velocities by
\[
\bv\sjk = \bv\sk - \bv\sj, \quad v\sjk = \vert \bv\sjk \vert , \quad j,k = 1, \ldots, N.
\]

Then with $d_j = \vert \bd\sj \vert$,
\begin{eqnarray}
\bv_{1,j} &=& v^2 (\bd\sj + \mu_{12} \hat{\bv} / (m_1 v)) \neq 0 \quad\textrm{ if } v > \mu_{12}/(m_1 d_j), \textrm{ } j = 3, \ldots, N, \nonumber \\
\bv_{2,j} &=& v^2 (\bd\sj - \mu_{12} \hat{\bv} / (m_2 v)) \neq 0 \quad\textrm{ if } v > \mu_{12}/(m_2 d_j), \textrm{ } j = 3, \ldots, N, \label{Art0eq 4.9 bis} \\
\bv_{j,k} &=& v^2 (\bd\sk - \bd\sj) \neq 0 \qquad\qquad\textrm{ } j,k = 3, \ldots, N. \nonumber
\end{eqnarray}

We denote $\hat{\bv}_{jk} = \bv\sjk / \vert \bv\sjk \vert.$ We assume for all pairs $(j,k)$ with $q_{j,k} \neq 0$ that $\vert \hat{\bv}\sjk \cdot \hat{\bE} \vert \leq \delta$ for some $0 \leq \delta < 1.$ It follows that in our high-velocity states the relative average velocity of the pair $(1,2)$ is $v$ while all other particles travel with minimal velocity proportional to $v^2$ relative to each other as well as with respect to particles 1 and 2.

The relative momentum of particles $j$ and $k$ is
\beq\label{Art1eq 3.7}
\bp\sjk = -i \nabla_{(\tbx\sk - \tbx\sj)},
\ene
where in the derivative the positions of all other particles, as well as of the center of mass, are kept fixed.
The relative velocity of the pair $(j,k)$ is
\beq\label{III}
\bp\sjk / \mu\sjk = \tilde\bp\sk / m\sk - \tilde\bp\sj / m\sj = \bp\sk / \mu\sk - \bp\sj / \mu\sj.
\ene

It follows from the definition that $\phi_0 \in \Sscr(\R^{n(N-1)})$ and that
\beq\label{Art1eq 3.4}
\Phi_{\bsv} = e^{i \mu_{12} \bsv \cdot \bsx} \prod_{j=3}^N e^{i \mu_{j} \bsv\sj \cdot \bsx\sj} \Phi_0.
\ene

Moreover, by \eqref{Art0eq 4.1}
 $$
| \tbx\sk - \tbx\sj | = | \bx\sk - \bx\sj | \leq | \bx\sk | + | \bx\sj |, \ j,k=1,\cdots , N, \ |\bx_1| \leq
|\bx|, | \bx_2 | \leq
|\bx|.
$$ Hence,
we have good initial localization uniformly in $\bv$,
\beq\label{Art1eq 3.5}
\Vert (1 + \vert \tilde{\bx}\sk - \tilde{\bx}\sj \vert^2)^{2} \Phi_{\bsv} \Vert \leq C, \quad j,k = 1, \ldots, N.
\ene
Additionally, by (1.15) there are functions $f\sjk \in C_0^\infty (B_{\mu\sjk\eta\sjk})$ such that
\beq\label{Art1eq 3.6}
\Phi_{\bsv} = f\sjk(\bp\sjk - \mu\sjk \bv\sjk) \Phi_{\bsv},
\ene
where $\mu_{jk}$ is the reduced mass of the pair $(j,k)$,
\beq\label{Art1eq 3.7b}
\mu\sjk = \frac{m_j m_k}{m_j + m_k}.
\ene

Furthermore, $\eta_{12} = \eta, \eta_{1j} = 2 (1 + \eta \mu_{12} / m_1), \eta_{2j} = 2 (1 + \eta \mu_{12} / m_2), j=3, \ldots, N,$ and $\eta_{jk}=4$, for $j,k = 3, \ldots, N$.

Note that by \eqref{Art0eq 4.1}
\beq \bx=\tbx\sdos -\tbx\suno = i\frac{\partial}{\partial\bp},\quad
\tbx_k-\tbx_j= i \frac{\partial}{\partial\bp_k} - i
\frac{\partial}{\partial \bp_j}, \quad j,k=3,\ldots, N, \label{Art0eq 4.14}
\ene

\beq
\tbx_k-\tbx\suno=
i\frac{\partial}{\partial \bp_k}
+ \frac{\mu_{12}}{m\suno} i
\frac{\partial}{\partial\bp},
\quad
\tbx_k - \tbx\sdos=
i\frac{\partial}{\partial \bp\sk}
- \frac{\mu_{12}}{m\sdos} i
\frac{\partial}{\partial\bp},
\quad
k=3,\ldots,N. \label{Art0eq 4.15}
\ene

As in Enss and Weder \cite{EnssWeder1995} and Weder \cite{Weder1996MISCIAEF}, \eqref{Art1eq 3.5}, \eqref{Art1eq 3.6},  \eqref{Art0eq 4.14} and \eqref{Art0eq 4.15} allow us to reduce the proofs in the N-body case to the ones for two bodies. We introduce below an appropriate class of potentials where $D^{\beta}, D^{\alpha_0}$ denotes the derivative with the usual multi-index notation.

\begin{definition}\label{Art2df 1.1}
We denote by $\Vscr_0$  the class of real-valued potentials, $V^0 (\bx)$, defined on $\R^{n}$ with values in $\R$ such that $V^0 (\bx) = V^{0,\,vs} (\bx) + V^{0,\,l} (\bx)$
with $V^{0,\,vs} (\bx) \in \Vscr_{0,\,vs},$ $V^{0,\,l} (\bx) \in \Vscr_{0,\,l},$
where $\Vscr_{0,\,vs}$ is the class of real-valued potentials, $V^{0,\,vs}$, that are relatively bounded with respect to the Laplacian with relative bound zero and
\beq\label{Art1eq 1.3 bis}
\int^\infty\scero d R\; \Big\| V^{0,\,vs}(\bx)\, (- \Delta + I)^{-1} F (|\bx| \geq R) \Big\| < \infty.
\ene
$\Vscr_{0,\,l}$ is the class of real-valued potentials $V^{0,\,l}$ that satisfy $V^{0,\,l}(\bx) \in C_\infty^1(\R^n),$ the space of all continuously differentiable functions that tend to zero at infinity, and that
\begin{eqnarray}
| D^{\beta} V^{0,\,l}(\bx)| &\leq& C(1+|\bx|)^{-\gamma_1}, \ \ \ | \beta | = 1, \gamma_1 > 3/2 \label{Art1eq 1.5 10},
\end{eqnarray} where without loss of generality we assume that $ \gamma_1 \leq 2$, otherwise $V^{0,\,l}$ would be of short range.
\end{definition}

Let $\epsilon_0$ satisfy:  $0< \epsilon_0 < \gamma_1-\frac{3}{2}$. After Hörmander \cite{Hormander1976}, we can write, without loss of generality that, for all $V^0 (\bx) \in \Vscr_0$, $V^0 (\bx) = V^{0,\,vs} (\bx)+ V^{0,\,l} (\bx)$ with $V^{0,\,vs} \in \Vscr_{0,\,vs}$, $V^{0,\,l} \in C^4(\R^n)$ and
\beq\label{Art1eq 1.5 20}
 | D^{\alpha_0} V^{0,\,l}(\bx)| \leq C(1+| \bx |)^{-1-|\alpha_0|(\epsilon_0+1/2)}, \textrm {  for } 2 \leq | \alpha_0 | \leq 4.
\ene

The more intuitive condition
$$
\int^\infty\scero d R\; \Big\| F (|\bx| \geq R) \, V^{0,\,vs}(\bx)\, (- \Delta + I)^{-1} \Big\| < \infty,
$$
by Reed and Simon \cite{ReedSimonIII}, is equivalent to the decay property \eqref{Art1eq 1.3 bis}.

\begin{definition}\label{Art2df 1.2} \cite{AdachiMaehara2007}.
We denote by $\Vscr_E$  the class of potentials, $V^E (\bx)$, defined on $\R^{n}$ with values in $\R$ such that $V^E (\bx) = V^{E,\,vs} (\bx) + V^{E,\,s} (\bx) + V^{E,\,l} (\bx)$ with $V^{E,\,vs} (\bx) \in \Vscr_{E,\,vs}$, $V^{E,\,s} (\bx) \in \Vscr_{E,\,s},$ and $V^{E,\,l} (\bx) \in \Vscr_{E,\,l},$
where $\Vscr_{E,\,vs}$ is the class of real-valued potentials, $V^{E,\,vs}$, that satisfy $V^{E,\,vs} = V_1^{E,\,vs} + V_2^{E,\,vs}$ with $(1+|x_1|)V_1^{E,\,vs}$ relatively bounded with respect to the Laplacian with relative bound zero and $V_2^{E,\,vs}$ bounded and that
\beq\label{Art1eq 1.3}
\int^\infty\scero d R\; \Big\| V^{E,\,vs}(\bx)\, (- \Delta + I)^{-1} F (|\bx| \geq R) \Big\| < \infty.
\ene
$\Vscr_{E,\,s}$ is the class of real-valued potentials $V^{E,\,s}$ that satisfy $V^{E,\,s}(\bx) \in C^1(\R^n)$ and that
\begin{eqnarray}
| V^{E,\,s}(\bx)| &\leq& C(1+|\bx|)^{-\gamma},\label{Art2eq 1.5} \\
| D^{\beta} V^{E,\,s}(\bx)| &\leq& C(1+|\bx|)^{-1-\alpha}, \ \ \ | \beta | = 1, \label{Art2eq 1.6}
\end{eqnarray}
with some $1/2 < \alpha \leq \gamma \leq 1$.
$\Vscr_{E,\,l}$ is the class of real-valued potentials $V^{E,\,l}$ that satisfy $V^{E,\,l}(\bx) \in C^2(\R^n)$ and that
\begin{eqnarray}
| D^{\beta} V^{E,\,l}(\bx)| &\leq& C(1+|\bx|)^{-\gamma_D-\mu| \beta |}, \ \ \ | \beta | \leq 2, \label{Art2eq 1.6 05}
\end{eqnarray}
with $0<\gamma_D \leq 1/2$ and $1-\gamma_D < \mu \leq 1.$
\end{definition}

The class of potentials $\Vscr_E$ in Definition \ref{Art2df 1.2} is the same as in Adachi and Maehara \cite{AdachiMaehara2007}. Again we can assume, without loss of generality by \cite{Hormander1976}, that for all $V^E (\bx) \in \Vscr_E$, $V^E (\bx) = V^{E,\,vs} (\bx)+ V^{E,\,s} (\bx) + V^{E,\,l} (\bx)$ with $V^{0,\,vs} \in \Vscr_{0,\,vs}$, $V^{E,\,s} (\bx) \in \Vscr_{E,\,s},$ $V^{E,\,l} \in C^4(\R^n)$ with $V^{E,\,l}$ satisfying \eqref{Art2eq 1.6 05} and
\beq
 | D^{\beta} V^{E,\,l}(\bx)| \leq C(1+| \bx |)^{-\gamma_D-\mu(2+| \beta |)/2}, \ \ \ 3 \leq | \beta | \leq 4. \label{Art2eq 1.6 10}
\ene

We call the potentials $V^{0,\,vs}$ and $V^{E,\,vs}$ very short-range, the potential $V^{E,\,s}$ short-range and the potentials $V^{0,\,l}$ and $V^{E,\,l}$ long-range.

For a particle with mass $m$ and charge $q$, there is a formula for the free time evolution, it was proven simultaneously by Avron and Herbst \cite{AvronHerbst1977} and by Veseli\'{c} and Weidmann \cite{VeselicWidmann1977}. There is also a generalization for the time-dependent case considered by Kitada and Yajima \cite{KitadaYajima1995},
\beq\label{Art1eq 2.1}
e^{-it(\bsp^2/(2m) - qE\bsx_1)} = e^{iqE\bsx_1 t} e^{-i t^3 q^2 E^2/(6m)} e^{-i\bsp_1 qEt^2/(2m)}e^{-i t\bsp^2/(2m)}.
\ene

We will also make frequent use of the following relations that are
obtained under translation in configuration or momentum space generated by $\bx$ or $\bp$, respectively,
\beq\label{Art1eq 2.2}
e^{i\bsp \cdot \bsv t} f(\bx) e^{-i\bsp \cdot \bsv t} = f(\bx + \bv t),
\ene
\beq\label{Art1eq 2.3}
e^{-im\bsv \cdot \bsx} f(\bp) e^{im\bsv \cdot \bsx} = f(\bp + m\bv),
\ene
for any measurable and bounded function f. In particular,
\eqref{Art1eq 2.3} implies that
\beq\label{Art1eq 2.4}
e^{-im\bsv \cdot \bsx} e^{-i t\bsp^2/(2m)} e^{im\bsv \cdot \bsx} = e^{-i\bsp \cdot \bsv t} e^{-i t\bsp^2/(2m)} e^{-im v^2 t/2},
\ene
where $v=|\bv|$. Since $ e^{i t\bsp^2/(2m)} \, \bx \, e^{-i t\bsp^2/(2m)} = \bx + t \, \bp/m $ and functional calculus, \beq\label{Art0eq 3.14 05}
e^{i t\bsp^2/(2m)} \, f (\bx) \, e^{-i t\bsp^2/(2m)} = f (\bx + t \, \bp/m).
\ene


We denote by $\be_1=(1,0,\ldots, 0)$ the unit vector along the $x_1$ direction
and $\hat{\bE}=\bE/\vert \bE \vert$. We designate by $q\sjk = (q\sk m\sj - q\sj m\sk)/(m\sj + m\sk)$ the relative charge of the pair $(j, k)$ and we denote by $\sum_{j < k}^0$ and $\sum_{j < k}^E$, respectively, the sum over all indices, $j < k, j,k = 1, \ldots, N$, with $q_{jk} = 0$, and $q_{jk} \neq 0$.

We assume that the potential of the N-body system is a multiplication operator that is a sum of pair potentials, \beq\label{Art1eq 1.21}
V = \sum_{j < k}^0 V\sjk^0 (\tbx\sk-\tbx\sj) + \sum_{j < k}^E V\sjk^E (\tbx\sk-\tbx\sj),
\ene
with $V_{jk}^0 \in \Vscr_0$ (see Definition \ref{Art2df 1.1}), and $V_{jk}^E \in \Vscr_E$ (see Definition \ref{Art2df 1.2}). By using a decomposition of $H\scero$ as in \eqref{Art1eq 1.15} for each pair $(j, k)$ we see that each of the
pair potentials $V\sjk^0$ and $V\sjk^E$ are relatively bounded with respect to $H\scero$ with relative bound zero. Note that for a given pair the corresponding pair potential belongs to $\Vscr_0$ if the relative charge of the pair is zero and that it belongs to $\Vscr_E$ if the relative charge is different from zero. Then V is relatively bounded with respect to $H\scero$ with relative bound zero and the interacting Hamiltonian,
\beq\label{Art1eq 1.22}
H = H_0 + V,
\ene
is self-adjoint on $D(H)=D(H\scero)$.

It is convenient to split the potential into the very short-, short- and long-range potentials.
For this purpose we define
\begin{eqnarray}\label{Art1eq 1.26}
\Vscr_{VSR}
&= & \left\{ V^{VS} = \sum_{j<k}^0 V_{jk}^{0,\,vs} (\tbx\sk - \tbx\sj) + \sum_{j<k}^E V_{jk}^{E,\,vs} (\tbx\sk - \tbx\sj)
\ \Big| \
V_{jk}^{0,\,vs} \in \Vscr_{0,\,vs}, \ V_{jk}^{E,\,vs} \in \Vscr_{E,\,vs}
\right\}, \phantom{ab}
\end{eqnarray}
\begin{eqnarray}\label{Art1eq 1.26b}
\Vscr_{SR}
&= & \left\{ V^{S} = \sum_{j<k}^E V_{jk}^{E,\,s} (\tbx\sk - \tbx\sj) \ \Big| \ V_{jk}^{E,\,s} \in \Vscr_{E,\,s} \right\},
\end{eqnarray}
\beq\label{Art1eq 1.27}
\Vscr_{LR} = \left\{ V^{L} = \sum_{j<k}^0 V_{jk}^{0,\,l} (\tbx\sk - \tbx\sj) + \sum_{j<k}^E V_{jk}^{E,\,l} (\tbx\sk - \tbx\sj) \ \Big| \ V_{jk}^{0,\,l} \in \Vscr_{0,\,l}, V_{jk}^{E,\,l} \in \Vscr_{E,\,l} \right\}.
\ene

Then
\beq\label{Art1eq 1.28}
V = V^{VS} + V^S + V^L,
\quad \quad H = H_0 + V = H_0 + V^{VS} + V^S +  V^L.
\ene

Let $S^D= S^D(V^L; V^{VS} + V^S)$ be the Dollard modified scattering operator defined in equation \eqref{Art1eq 1.29} below.

Our main results are the reconstruction formulae given in Theorems \ref{Art2tr 2.1} and \ref{Art1tr 3.1}
that we prove in Section \ref{Reconstruction Formulae}. The uniqueness result given in Theorem \ref{Art1te 1.2} follows from Theorem \ref{Art2tr 2.1}.

\begin{theorem}\label{Art1te 1.2} Let $\gamma_1$ be as in Definition \ref{Art2df 1.1} and, $\gamma_D$ and $\mu$ as in Definition \ref{Art2df 1.2}. If there are two pairs $1 \leq j < k \leq N,$ $1 \leq j' < k' \leq N,$ with $q\sjk \neq 0$ and $q\sjpkp=0
$
we assume that $\gamma_1 > 3 - 4 (\gamma_D + \mu)/3.$ Then,
\begin{enumerate}
\item Suppose that $V^i = V^{VS, \, i}+V^{S, \, i}+V^{L, \, i} \in \Vscr_{VSR}+\Vscr_{SR}+\Vscr_{LR}, \, i=1,2,$ and that $S^D(V^{L, \, 1}; V^{VS, \, 1} + V^{S, \, 1}) = S^D(V^{L, \, 2}; V^{VS, \, 2} + V^{S, \, 2}).$ Then, $V\euno = V\edos.$
\item Furthermore, it is possible to uniquely reconstruct the total potential $V$ from any Dollard scattering operator $S^{D}.$
\end{enumerate}
\end{theorem}


\begin{remark} {\rm
Note that in item 1 of Theorem \ref{Art1te 1.2} it is enough to assume that the high-velocity limits of $S^D(V^{L, \, 1}; V^{VS, \, 1} + V^{S, \, 1})$ and  $S^D(V^{L, \, 2}; V^{VS, \, 2} + V^{S, \, 2})$ are the same. Furthermore, we prove item 2 of Theorem \ref{Art1te 1.2} giving a method for the unique reconstruction of $V$ from the high-velocity limit of any Dollard scattering operator. See the reconstruction formulae \eqref{Art1eq 3.24}, \eqref{Art1eq 3.23} and the proof of Theorem \ref{Art1te 1.2}.

}
\end{remark}

\begin{remark} {\rm
For a given $V^L \in \Vscr_{LR}$ let us define, as in \cite{AdachiMaehara2007} and \cite{Weder1996MISCIAEF}, the scattering map $S\suno := S^{D}(V^L; \cdot ),$ $S\suno (Q) = S^D (V^L; Q), Q \in \Vscr_{VSR} + \Vscr_{SR},$ an operator from $\Vscr_{VSR} + \Vscr_{SR}$ into the Banach space $\Lscr(\Hscr)$ of all bounded operators in $\Hscr.$ Clearly, Theorem \ref{Art1te 1.2} implies that $S\suno = S^{D}(V^L; \cdot )$ is injective.
}
\end{remark}

\begin{remark}\label{RemarkInjectivity3} {\rm
For a given $V^L \in \Vscr_{LR}$ and a given $V^S \in \Vscr_{SR}$ we define the scattering map $S\sdos := S^{D}(V^L; \cdot + V^S),$ $S\sdos(V^{VS}) = S\sdos (V^L; V^{VS} + V^{S}),$ an operator from $\Vscr_{VSR}$ into $\Lscr(\Hscr).$ It is immediate that Theorem \ref{Art1te 1.2} implies that $S\sdos = S^{D}(V^L; \cdot + V^S)$ is injective. However, as we show in Remark \ref{RemarkArt1tr 3.1} this result can also be proven using the reconstruction formula \eqref{Art1eq 3.23} given in Theorem \ref{Art1tr 3.1}, that is simpler than the formula \eqref{Art1eq 3.24} in Theorem \ref{Art2tr 2.1}, because in \eqref{Art1eq 3.23} it is not necessary to take the commutator of $S^{D}$ with a component of the momentum operator. This is important in applications where the tail at infinity of the potential is already known and one wishes to uniquely reconstruct $V^{VS}$ assuming that $V^S$ and $V^L$ are known.}

\end{remark}

\begin{remark}\label{Reviewer20120615_03}{\rm
Under the Stark effect, for a pair potential where the relative charge is not zero, the short-range decay rate at infinity of this potential depends on $\gamma$ given in our equation \eqref{Art2eq 1.5}. Theorem \ref{Art1te 1.2} is proved by the first time by Weder \cite{Weder1996MISCIAEF}, where he considers $\gamma > 3/4$ and N-Body pair potentials which are short-range if the corresponding relative charge is not zero and long-range if the corresponding relative charge is zero. Then, for two body short-range potentials, Nicoleau \cite{Nicoleau2003} proves this Theorem with $\gamma > 1/2,$ the dimension of the space $n\geq3$ and the regularity and decay
of the potential:
\beq\label{NicoleauPotentials}
V:  \quad |\partial_x^{\beta} V(x)| \leq C_{\beta} (1+|x|)^{-\gamma-|\beta|},
\ene
for all multi-index $\beta.$ Later, in the two-body case, Adachi and Maehara \cite{AdachiMaehara2007} improve the results of Nicoleau \cite{Nicoleau2003} because, besides $\gamma > 1/2,$ they relax the conditions on the derivatives on the potential and use dimension $n \geq 2.$ Furthermore, Adachi and Maehara \cite{AdachiMaehara2007} consider long-range potentials whereas Nicoleau \cite{Nicoleau2003} does not.

We improve the N-body results of Weder \cite{Weder1996MISCIAEF}.
Our potential $V$ is given, by
$$
V = \sum_{j<k}^0 \left( V_{jk}^{0,\,vs} (\tbx\sk - \tbx\sj) + V_{jk}^{0,\,l} (\tbx\sk - \tbx\sj) \right) + \sum_{j<k}^E \left( V_{jk}^{E,\,vs} (\tbx\sk - \tbx\sj) + V_{jk}^{E,\,s} (\tbx\sk - \tbx\sj) + V_{jk}^{E,\,l} (\tbx\sk - \tbx\sj) \right)
$$
where, for all $1 \leq j<k \leq N,$ $V_{jk}^{0,\,vs} \in \Vscr_{0,\,vs},$ $V_{jk}^{0,\,l} \in \Vscr_{0,\,l},$ $V_{jk}^{E,\,vs} \in \Vscr_{E,\,vs},$ $V_{jk}^{E,\,s} \in \Vscr_{E,\,s},$ $V_{jk}^{E,\,l} \in \Vscr_{E,\,l}.$
Our potential $V\sjk^{E,\,l}$ has no counterpart in \cite{Weder1996MISCIAEF}, i.e. potentials that are long-range with respect to the Stark effect, when the relative charge $q\sjk \neq 0,$ are not allowed in \cite{Weder1996MISCIAEF} whereas, here, we do. This is our first improvement over \cite{Weder1996MISCIAEF}. Secondly, in equation (1.4) of \cite{Weder1996MISCIAEF} $\gamma > 3/4$ and in our equation \eqref{Art2eq 1.5} we have $\gamma > 1/2,$ thus we improve the results of \cite{Weder1996MISCIAEF} because our potential $V\sjk^{E,\,s}$ is allowed to have the optimal short-range decay rate at infinity.


We give a reconstruction formula with an error term that goes to zero as an inverse power of the velocity, that depends on the decay rate of the potentials, see Theorems \ref{Art2tr 2.1} and \ref{Art1tr 3.1}. If we only assume \eqref{Art1eq 1.3}, our results coincide with those of Adachi and Maehara \cite{AdachiMaehara2007}, in the case $N=2$ and $q_{12} \neq 0.$ If, instead of \eqref{Art1eq 1.3}, we assume \eqref{Art1eq 2.5}, we give a sharper error term than theirs. In this sense, we can say that we obtain a new result, even in the two body case.} \hfill \bull

\end{remark}

In this paper, we prove Theorem \ref{Art1te 1.2} by extending to the N-body case the results obtained, in the
two-body case,
by Adachi and Maehara \cite{AdachiMaehara2007} using the
the findings published in 1993 \cite{EnssWeder1993}, 1994 \cite{EnssWeder1994}, 1995 \cite{EnssWeder1995} by V. Enss and R. Weder where a new time-dependent method was developed. Here, physical propagation properties of finite energy wave functions are used to estimate the high-velocity behavior of solutions of the Schrödinger equation and solve inverse scattering problems in quantum mechanics. It is intuitive from the point of view of the physics related to the problem. Contrary to the stationary approach, this method can be applied to study non-linear equations \cite{SandWeder2006, Weder1999, Weder2000, Weder2000bis, Weder2000iv, Weder2001, Weder2001bis, Weder2001tris, Weder2002bis, Weder2003bis, Weder2005, Weder2005bis}. Lately, this time-dependent approach has been exploited to study: Hamiltonians with electric and magnetic fields \cite{Arians1996, Arians1997, Arians1998, Jung2005}, N-body systems \cite{EnssWeder1994, EnssWeder1995, EnssWeder1996, Weder1996MISCIAEF, Weder1996}, the Stark effect \cite{Weder1996MISCIAEF, AdachiMaehara2007, Nicoleau2003, Nicoleau2005, AdachiEtAl2011}, the Aharanov-Bohm effect \cite{Nicoleau2000, Weder2002, BallesWeder2008, BallesWeder2009, BallesWeder2011,Weder2011-0, Weder2011}, time-dependent potentials \cite{Weder1996, Nicoleau2005}, Dirac equation \cite{Ito1998, Ito1998bis, Jung1996, Jung1997, Enss1998}, Klein-Gordon equation \cite{EnssJung1999, Enss1998, Weder2000iv, Weder2002bis}, mass and charge of black holes \cite{DaudeNicoleau2008, DaudeNicoleau2010}, amongst others.

\section{Reconstruction Formulae}\label{Reconstruction Formulae}\sss

Let us define
\beq\label{Art1eq 1.23 -25}
V^{vs}\sjk  =
\begin{cases}
V^{0,\,vs}_{jk}, & \textrm{if } q\sjk = 0, \\
V^{E,\,vs}_{jk}, & \textrm{if } q\sjk \neq 0,
\end{cases}
\ \
V^{s}\sjk  =
\begin{cases}
0, & \textrm{if } q\sjk = 0, \\
V^{E,\,s}_{jk}, & \textrm{if } q\sjk \neq 0,
\end{cases}
\ \
V^{l}\sjk  =
\begin{cases}
V^{0,\,l}_{jk}, & \textrm{if } q\sjk = 0, \\
V^{E,\,l}_{jk}, & \textrm{if } q\sjk \neq 0.
\end{cases}
\ene where $V^{0,\,vs}_{jk}$ and $V^{0,\,l}_{jk}$ are defined in Definition \ref{Art2df 1.1}, and $V^{E,\,vs}_{jk}, V^{E,\,s}_{jk}$ and $V^{E,\,l}_{jk}$ are defined in Definition \ref{Art2df 1.2}. Moreover, the decays of $V^{0,\,l}_{jk}$ and $V^{E,\,l}_{jk}$ are to be taken as in \eqref{Art1eq 1.5 20}, \eqref{Art2eq 1.6 05} and \eqref{Art2eq 1.6 10}, respectively.

We introduce the Graf-type modifier \eqref{Art1eq 1.23 -05} Graf \cite{Graf1991} and Zorbas \cite{Zorbas1978}, to define auxiliary wave operators, whose existence and completeness were proven in the N body case for long-range Stark Hamiltonians by Adachi and Tamura \cite{AdachiTamura1995}. We note that the $\bv$ dependence of the Graf-type modifier \eqref{Art1eq 1.23 -05} is first introduced in Adachi and Maehara \cite{AdachiMaehara2007} by taking into account the $\bv$ dependence of $\Phi_{\bsv}.$ In Graf \cite{Graf1991} and Zorbas \cite{Zorbas1978}, $\bv$ is taken $0$ in the definiton of the Graf-type modified wave operators.
We define the Graf-type modifier \cite{AdachiMaehara2007}, \cite{Graf1991}, \cite{Zorbas1978} and the Dollard-type modifier by \eqref{Art1eq 1.23 -05}, and \eqref{Art1eq 1.24}, respectively
\begin{eqnarray}
\tilde{U}_{G,v} (t) &=& \exp \left(-i \sum_{j<k}^E \int_0^t ds \, V^{E,\,s}_{jk}(\bv\sjk \, s + \be_1 q\sjk E s^2/(2\mu\sjk) ) \right), \label{Art1eq 1.23 -05}
\\
\tilde{U}_{D}(t) &=& \exp \left(-i \sum_{j<k} \int_0^t ds \, V_{jk}^{l}(s \bp\sjk/\mu\sjk + \be_1 q\sjk E s^2/(2\mu\sjk)) \right). \label{Art1eq 1.24}
\end{eqnarray}

For completeness we mention, for the short-range case, the modified
Graf
propagator, the modified
Graf
wave operators \cite{Graf1991}, \cite{Zorbas1978} and the
wave operators for the free channel which are defined, respectively, by \eqref{Art1eq 1.25 -05 sr} \eqref{Art1eq 1.25 sr} and \eqref{Art1eq 1.25 05 sr}:
\begin{eqnarray}
U^{G,v}(t) &=& e^{-itH\scero} \tilde{U}_{G,v}(t), \label{Art1eq 1.25 -05 sr} \\
\Omega_\pm^{G,v} &=& \textrm{s}-\lim\limits_{t\to \pm \infty} e^{itH}\; U^{G,v}(t), \label{Art1eq 1.25 sr} \\
W_\pm &=& \textrm{s}-\lim\limits_{t\to \pm \infty} e^{itH}\; e^{-itH\scero}. \label{Art1eq 1.25 05 sr}
\end{eqnarray}

The modified Dollard-Graf
propagator, the modified
Dollard-Graf
wave operators \cite{Graf1991}, \cite{Zorbas1978} and the
modified Dollard
wave operators for the free channel are defined, respectively, by \eqref{Art1eq 1.25 -05} \eqref{Art1eq 1.25} and \eqref{Art1eq 1.25 05}:
\begin{eqnarray}
U^{D,G,v}(t) &=& e^{-itH\scero} \tilde{U}_{D}(t) \tilde{U}_{G,v}(t), \label{Art1eq 1.25 -05} \\
\Omega_\pm^{D,G,v} &=& s-\lim\limits_{t\to \pm \infty} e^{itH}\; U^{D,G,v}(t), \label{Art1eq 1.25} \\
W_\pm^{D} &=& s-\lim\limits_{t\to \pm \infty} e^{itH}\; e^{-itH\scero} \tilde{U}_{D}(t). \label{Art1eq 1.25 05}
\end{eqnarray}

Tamura proved the existence of the $W_\pm$ for short range N-Body Stark systems \cite{Tamura1993}, Korotyaev \cite{Korotyaev1988} did it for the case N=3. Adachi and Tamura \cite{AdachiTamura1995}, and, Herbst, Møller and Skibsted \cite{HerbstMollerSkibsted1996} proved the existence of $W_\pm^{D}$ given by \eqref{Art1eq 1.25 05} for the N-Body long-range case. Actually, the existence of the $W_\pm^{D}$ and $W_\pm^{}$ also follows from our estimates. We give the simple proof of the existence of $\Omega_\pm^{D,G,v}$ and $\Omega_\pm^{G,v}$ in Proposition \ref{DollardGrafWaveOperators}. The Dollard scattering operator $S^{D}$ from the free channel to the free channel is defined as
\beq\label{Art1eq 1.29}
S^{D}= S^{D}(V^L; V^{VS} + V^S) := (W_+^{D})^* W_-^{D},
\ene
$S^{D}$ is not unique because there is more than one short- and long-range splitting of the potential. We also mention the scattering operator $S$ from the free channel to the free channel defined for the short-range case as
\beq\label{Art1eq 1.29 sr}
S^{} = (W_+^{})^* W_-^{}.
\ene

Proposition \ref{Art0pr 2.1}, below, shall be frequently used in this text. Its proof is given in the Proposition 2.10 in Enss \cite{Enss1983}.
\begin{prop}\label{Art0pr 2.1}
For any $f \in C_0^\infty(\Rn)$ with $\hbox{\rm{supp }} f \subset B_{m\eta_0},$ for some $m,\eta_0>0$ and any $l=1,2,3,\ldots$ there is a constant $C_l$ such that the following estimate is true:
\[
\left\Vert F(\bx \in \mathcal{M}') \, e^{-it\bsp^2/(2m)} f(\bp-m\bv) F(\bx \in \mathcal{M}) \right\Vert \leq C_l (1+r+|t|)^{-l},
\]
for every $\bv \in \Rn, \, t \in \R$ and any measurable sets $\mathcal{M}', \, \mathcal{M}$ such that $r:= \textrm{dist}(\mathcal{M}', \, \mathcal{M}+\bv t) - \eta_0 |t| > 0.$
\end{prop}

To treat the case whether or not the relative charge $q\sjk$ is zero, we define
\beq\label{Art1eq 2.13 c}
\delta\sjk :=
\begin{cases}
\delta , & \textrm{if } q\sjk \neq 0, \\
0 , & \textrm{if } q\sjk = 0.
\end{cases}
\ene where $\delta$ is such that $\vert \hat{\bv}\sjk \cdot \hat{\bE} \vert \leq \delta < 1,$ for all integers $1 \leq j < k \leq N$ with $ q_{jk} \neq 0.$

A cornerstone throughout this work is the existence of $0 < \delta_1, \delta_2 \leq 1$ such that
\beq\label{Art1eq 2.13 b}
\vert \bv t + \be_1 q_{12}Et^2/(2\mu_{12}) \vert \geq \sqrt{\delta_1 \vert vt \vert^2 + \delta_2 (q_{12}E/(2\mu_{12}))^2 t^4} \geq \sqrt{\delta_1} \vert vt \vert.
\ene
When $q_{12} = 0$, we can take $\delta_1=\delta_2=1,$ and if $q_{12} \neq 0,$ we use $\delta_1=\delta_2=1-\delta,$
Moreover, if $0 \leq \tilde\sigma \leq 1, \, q_{12} \neq 0, \, | \bp | \leq \mu_{12} \eta,$ and $ \eta / v < \sqrt{1-\delta}/4,$ from a simple computation, there exist two positive constants $c_1$ and $c_2$ such that
$$
\vert t \, \bp/\mu_{12} + \bv t + \be_1 q_{12}Et^2/(2\mu_{12}) \vert \geq c_1 \vert vt \vert,
$$
$$
\vert t \, \bp/\mu_{12} + \bv t + \be_1 q_{12}Et^2/(2\mu_{12}) \vert \geq c_2 t^2,
$$
\beq\label{Art1eq 2.13 b3}
\vert t \, \bp/\mu_{12} + \bv t + \be_1 q_{12}Et^2/(2\mu_{12}) \vert \geq c_1^{\tilde\sigma} c_2^{1-\tilde\sigma} \vert vt \vert^{\tilde\sigma} t^{2(1-\tilde\sigma)}.
\ene

For any pair $(j,k),$ we establish three conditions:
$\zeta\sjk^{a}$ as $``\gamma_1<2$ and there is, at least, one pair $(j',k')$ with $q\sjpkp=0, \, V_{\jp,\kp}^{l} \neq 0,$ and either $\jp = j \textrm{ or } \jp = k \textrm{ or } \kp = j \textrm{ or } \kp = k \textrm{ or } \jp + j = 3",$ $\zeta\sjk^{b}$ as $``\gamma_1=2$ and there is, at least, one pair $(j',k')$ with $q\sjpkp=0, \, V_{\jp,\kp}^{l} \neq 0,$ and either $\jp = j \textrm{ or } \jp = k \textrm{ or } \kp = j \textrm{ or } \kp = k \textrm{ or } \jp + j = 3",$ and $\zeta\sjk^{c}$ as ``there is no pair $(j',k')$ with $q\sjpkp=0, \, V_{\jp,\kp}^{l} \neq 0,$ and either $\jp = j \textrm{ or } \jp = k \textrm{ or } \kp = j \textrm{ or } \kp = k \textrm{ or } \jp + j = 3".$ We define the following constant, for any $\epsilon > 0$:
\beq\label{Art2eq 3.7 -10}
\theta\sjk :=
\begin{cases}
2-\gamma_1, & \hbox{\rm{if }} \zeta\sjk^{a}, \\
\epsilon, & \hbox{\rm{if }} \zeta\sjk^{b}, \\
0, & \hbox{\rm{if }} \zeta\sjk^{c}.
\end{cases}
\ene

\begin{lemma}\label{Art2le 3.1}
Let $\tilde{U}_{D}(t)$ be given as in \eqref{Art1eq 1.24}.
Then there exists a constant $C$, such that for all $t \in \R,$ for every $\bv\sjk \in \R^n,$ as in \eqref{Art0eq 4.9 bis}, with $v\sjk \geq 4\eta\sjk/\sqrt{1-\delta\sjk}$
and $ v=v_{12},$
for all $f\sjk \in  C_0^\infty (B_{\mu\sjk\eta\sjk}),$
for all integers $1 \leq j < k \leq N,$
$\textrm{  for all } \kappa > 0 \textrm{ and } 0 < \tilde\epsilon < \min \{4\epsilon_0, \, 2\gamma_D + 5\mu + \epsilon_0 - 5/2, \, 2\gamma_D + 6\mu - 3 \},$ one has that

\begin{eqnarray}
A\sjk &:=& \left\Vert \left(\tilde\bx\sk - \tilde\bx\sj\right) \: \tilde{U}_{D}(t) \prod_{j^\prime <
k^\prime} f_{j^\prime k^\prime} (\bp_{\jp\kp} - \mu\sjpkp \bv\sjpkp)
(1 + |\tilde\bx\sk - \tilde\bx\sj|\edos)^{-1/2}  \right\Vert \nonumber \\
&\leq& C
\begin{cases}
1 + v\sjk^{-(2-\gamma_1 /2)}|v\sjk t|^{2-\gamma_1}, & \hbox{\rm{if }} \zeta\sjk^{a}
, \\
1+v\sjk^{-1} \ln (1 +  v\sjk^{-1/2}| v\sjk t|), & \hbox{\rm{if }} \zeta\sjk^{b}
, \\
1, & \hbox{\rm{if }} \zeta\sjk^{c}
,
\end{cases}
\quad \leq \quad
C \left( 1 + v\sjk^{-(2-\gamma_1/2)}|v\sjk t|^{\theta\sjk}  \right),  \label{Art2eq 3.2b} \\
%
B\sjk &:=& \left\Vert F(|\tilde\bx\sk - \tilde\bx\sj| > \kappa | v\sjk t |) \, \tilde{U}_{D}(t) \prod_{j^\prime <
k^\prime} f_{j^\prime k^\prime} (\bp_{\jp\kp} - \mu\sjpkp \bv\sjpkp
)
(1 + |\tilde\bx\sk - \tilde\bx\sj|\edos)^{-2}  \right\Vert
\nonumber \\
&\leq & C \left(1 + \vert v\sjk t \vert \right)^{-2-\tilde\epsilon}.
\label{Art2eq 3.3}
\end{eqnarray}
\end{lemma}

\begin{dem}

By \eqref{Art0eq 4.14} and \eqref{Art0eq 4.15}, for $1 \leq a \leq 4,$ multiplication by $\left(\tilde\bx\sk - \tilde\bx\sj\right)^a$ becomes derivatives in the $\bp, \bp\sk, k=3,\ldots,N$ variables.

The norm
$$
\left\Vert \left(\tilde\bx\sk - \tilde\bx\sj\right)^a \: \tilde{U}_{D}(t) \prod_{j^\prime <
k^\prime} f_{j^\prime k^\prime} (\bp_{\jp\kp} - \mu\sjpkp \bv\sjpkp)(1 + |\tilde\bx\sk - \tilde\bx\sj|\edos)^{-a/2} \right\Vert
$$ is bounded by a finite sum of terms of the form $C \prod_{b=1}^{a} I_{\beta_b},$
with $I_{\beta_b}=1$ if the multi-index $\beta_b=0$ and
if
$|\beta_b|>0,$
\beq\label{Art2eq 3.3 beta_b 01}
I_{\beta_b} = \left\Vert \int_0^t |s|^{|\beta_b|} \left(D^{\beta_b} V_{\jp\kp}^{l}\right)\left(s(\bp_{\jp\kp}/\mu_{\jp\kp} + \bv\sjpkp) + \be_1 q_{\jp\kp} E s^2/(2\mu_{\jp\kp})\right) ds \: g_{j^\prime k^\prime} (\bp_{\jp\kp}) \right\Vert,
\ene
where $(j', k')$ is a pair of integers $1 \leq j' < k' \leq N$ such that $\jp = j \textrm{ or } \jp = k \textrm{ or } \kp = j \textrm{ or } \kp = k \textrm{ or } \jp + j = 3,$ $g\sjpkp \in  C_0^\infty (B_{\mu\sjpkp\eta\sjpkp})$ and $g\sjpkp = 1$ in the support of $f\sjpkp.$ Note that $\sum_{b=1}^{a} | \beta_b | \leq a.$

Below, we take $\tilde\sigma = 0,$ if $q_{\jp\kp} \neq 0$ and $\tilde\sigma=1,$ if $q_{\jp\kp}=0.$ We define
\beq\label{Art2eq 3.3 beta_b 05}
i_{\beta_b,v\sjpkp} (s)
:=
  \begin{cases}
    (1 +  |v\sjpkp s|)^{-\gamma_1}, & \textrm{if } q_{\jp\kp}=0 \textrm{ and }  |\beta_b | = 1, \cr
    (1 + |v\sjpkp s|)^{-1-|\beta_b|(\epsilon_0 + 1/2)}, & \textrm{if } q_{\jp\kp}=0 \textrm{ and } 2 \leq | \beta_b | \leq 4, \cr
    (1 + |s|^{2})^{-\gamma_D - \mu |\beta_b|}, & \textrm{if } q_{\jp\kp} \neq 0 \textrm{ and }  1 \leq | \beta_b | \leq 2, \cr
    (1 + |s|^{2})^{-\gamma_D - \mu (2+|\beta_b|)/2}, & \textrm{if } q_{\jp\kp} \neq 0 \textrm{ and } 3 \leq | \beta_b | \leq 4,
  \end{cases} \phantom{ab}
\ene
It follows from \eqref{Art1eq 1.5 10}, \eqref{Art1eq 1.5 20}, \eqref{Art2eq 1.6 05}, \eqref{Art2eq 1.6 10}, \eqref{Art1eq 2.13 b}, \eqref{Art1eq 2.13 b3}, \eqref{Art2eq 3.3 beta_b 01} and \eqref{Art2eq 3.3 beta_b 05}  that
\begin{eqnarray}
I_{\beta_b}
&\leq&
C \int_0^{|t|} s^{|\beta_b|} i_{\beta_b,v\sjpkp} (s) \, ds
\leq
C v\sjpkp^{-(|\beta_b|+1)\tilde\sigma/(2-\tilde\sigma)}\int_0^{v\sjpkp^{\tilde\sigma/(2-\tilde\sigma)}|t|} \tau^{|\beta_b|} i_{\beta_b,1} (\tau) \, d\tau.
\label{Weder 20110710 1.15 and 1.18}
\end{eqnarray}
Let us assume $|{\beta_b}|=1$ in \eqref{Weder 20110710 1.15 and 1.18}. If $q\sjpkp \neq 0,$
$$
I_{\beta_b}
\leq C \int_0^{|t|} \tau (1 +  \tau)^{2(-\gamma_D - \mu)} \, d\tau
\leq C.
$$
If $q\sjpkp=0
,$ we have that
\begin{eqnarray*}
I_{\beta_b}
&
\leq
&
C v\sjpkp^{-2} \int_0^{v\sjpkp |t|} \tau (1 +  \tau)^{-\gamma_1} \, d\tau
\leq
C v\sjpkp^{-2}
  \begin{cases}
    (1 + |v\sjpkp t|)^{2-\gamma_1}, & \textrm{if } \gamma_1 < 2, \cr
    \ln (1 +  |v\sjpkp t|), & \textrm{if } \gamma_1 = 2.
  \end{cases} \label{Art2eq 3.3 beta_b 01 -60}
\end{eqnarray*}
\begin{eqnarray*}
&
\leq
&
  C
  \begin{cases}
    1 + v\sjk^{-2} |v\sjk t|^{2-\gamma_1}, & \textrm{if } \gamma_1 < 2 \textrm{ and either } (j',k') = (j,k)  = (1,2), \cr & \textrm{or } (j',k') \neq (1,2) \textrm { and } (j,k) \neq (1,2),  \cr
    1 + v\sjk^{-(2+\gamma_1)}|v\sjk t|^{2-\gamma_1}, & \textrm{if } \gamma_1 < 2, (j',k') \neq (1,2) \textrm{ and } (j,k) = (1,2), \cr
    1 + v\sjk^{-(2-\gamma_1 /2)}|v\sjk t|^{2-\gamma_1}, & \textrm{if } \gamma_1 < 2, (j',k') = (1,2) \textrm{ and } (j,k) \neq (1,2) \cr
    v\sjk^{-2} \ln (1 +  |v\sjk       t|), & \textrm{if } \gamma_1 = 2 \textrm{ and either } (j',k') = (j,k)  = (1,2), \cr & \textrm{or } (j',k') \neq (1,2) \textrm { and } (j,k) \neq (1,2),  \cr
    v\sjk^{-4} \ln (1 +  |v\sjk^2       t|), & \textrm{if } \gamma_1 = 2, (j',k') \neq (1,2) \textrm{ and } (j,k) = (1,2), \cr
    v\sjk^{-1} \ln (1 +  |v\sjk^{1/2} t|), & \textrm{if } \gamma_1 = 2, (j',k') = (1,2) \textrm{ and } (j,k) \neq (1,2).
  \end{cases}
\end{eqnarray*}
This implies that \eqref{Art2eq 3.2b} is true.

In the other hand, similarly to \eqref{Weder 20110710 1.15 and 1.18}, we have that,
\begin{eqnarray}
I_{\beta_b} &\leq& C \int_0^{|t|} s^{|\beta_b|} i_{\beta_b,v} (s) \, ds
\leq C v^{-(|\beta_b|+1)\tilde\sigma/(2-\tilde\sigma)} \int_0^{v^{\tilde\sigma/(2-\tilde\sigma)}|t|} \tau^{|\beta_b|} i_{\beta_b,1} (\tau) \, d\tau \nonumber \\
&
\leq
&
C
  \begin{cases}
    1 + |v t|^{|\beta_b|(-\epsilon_0 + 1/2)}, & \textrm{if } q_{\jp\kp}=0 \textrm{ and }  1 \leq |\beta_b | \leq 4, \cr
    1, & \textrm{if } q_{\jp\kp} \neq 0 \textrm{ and }  1 \leq | \beta_b | \leq 2, \cr
    1 + |v t|^{\max \{\vert \beta_b \vert + 1 - 2\gamma_D - (\vert \beta_b \vert + 2)\mu, \, 0\}}, & \textrm{if } q_{\jp\kp} \neq 0 \textrm{ and } 3 \leq |\beta_b | \leq 4.
  \end{cases}
\nonumber
\end{eqnarray}
Then,
it follows that
$$
C \prod_{b=1}^{4} I_{\beta_b}
\leq
C (1+|vt|)^{2-\tilde\epsilon},
$$
hence
\begin{eqnarray*}
\kappa^4 |v\sjk t|^{4} \, B\sjk  &\leq& \left\Vert |\tilde\bx\sk - \tilde\bx\sj|^4 F(|\tilde\bx\sk - \tilde\bx\sj| > \kappa | v\sjk t |) \: \tilde{U}_{D}(t) \prod_{j^\prime < k^\prime} f_{j^\prime k^\prime} (\bp_{\jp\kp} - \mu\sjpkp \bv\sjpkp)(1 + |\tilde\bx\sk - \tilde\bx\sj|\edos)^{-2} \right\Vert  \\
& \leq & C (1+|vt|)^{2-\tilde\epsilon} \leq C (1+|v\sjk t|)^{2-\tilde\epsilon}.
\end{eqnarray*}
This proves Equation \eqref{Art2eq 3.3}.
\end{dem}

Lemma (\ref{Art2le 2.1}), below, is a generalization of equations (3.8) and (3.17) in Weder \cite{Weder1996MISCIAEF}. Note that conditions \eqref{Art1eq 1.3 bis} and \eqref{Art1eq 1.3} imply that
\[
\Vert V^{vs}(\bx) g(\bp) F(\vert \bx \vert \geq R) \Vert
\]
is an integrable function of R for all $g \in C_0^\infty(\R^n)$ (see Corollary 2.4 in Enss \cite{Enss1983}). It follows that potentials in $\Vscr_{E,vs}$ and $\Vscr_{0,vs}$, satisfy condition \eqref{Art1eq 2.5} below with $\rho = 0$. Of course, larger $\rho$ means faster decay.

\begin{lemma}\label{Art2le 2.1} Suppose that $V^{vs}\sjk$ is given as in \eqref{Art1eq 1.23 -25} and satisfies
\beq\label{Art1eq 2.5}
(1+R)^\rho \Vert V^{vs}\sjk(\tbx\sk - \tbx\sj) g(\bp\sjk) F (\vert \tbx\sk - \tbx\sj \vert \geq R) \Vert \in L^1((0, \infty), dR),
\ene
for some $0 \leq \rho \leq 1 $ and all $g \in C_0^\infty (\R^n)$, $\tilde{U}_{D}(t)$ is given as in \eqref{Art1eq 1.24}. Then, for all functions $f_{j^\prime k^\prime} \in C_0^\infty (B_{\mu\sjpkp \eta\sjpkp})$ with $1 \leq j^\prime < k^\prime \leq N,$ there is a function $h\sjk$ with $(1+\tau)^{\rho}h\sjk(\tau) \in L\euno ((0,\infty))$ such that for every $\bv\sjk \in \R^n$ with $v\sjk > c$ for some constant $0<c,$ we have the following estimate, for all integers $1 \leq j < k \leq N$:
\beq\label{Art1eq 3.9}
D\sjk
:=
\left\Vert V\sjk^{vs} (\tbx\sk - \tbx\sj) \, e^{-itH\scero} \tilde{U}_{D}(t) \prod_{j^\prime<k^\prime} f_{j^\prime k^\prime} (\bp_{\jp\kp} - \mu\sjpkp \bv\sjpkp) (1 + \vert  \tbx\sk - \tbx\sj \vert^2)^{-2} \right\Vert 
\leq
h\sjk (\vert v\sjk t \vert).
\ene
%
%
%
\end{lemma}

\begin{dem}
Let us take $g\sjk \in C_0^\infty (B_{\mu\sjk\eta\sjk})$ that satisfies $g\sjk \equiv 1$ on the support of $f\sjk.$
\begin{eqnarray}
D\sjk &\leq& I_1 + I_2 + I_3 \label{Art1eq 2.8 -05},
\end{eqnarray} where, for any positive constant $\lambda$,
\begin{eqnarray*}
I_1 &=& \left\Vert V\sjk^{vs} (\tbx\sk - \tbx\sj) g\sjk (\bp\sjk - \mu\sjk \bv\sjk) \right\Vert \, \left\Vert F(|\tilde\bx\sk - \tilde\bx\sj - \bv\sjk t - \be_1 q\sjk Et^2/(2\mu\sjk) | \geq \lambda | v\sjk t | 5/8) \, e^{-itH\scero} \right. \nonumber \\
& & \ \ \left. \times g\sjk (\bp\sjk - \mu\sjk \bv\sjk)  F(|\tilde\bx\sk - \tilde\bx\sj | < \lambda | v\sjk t |/8) \right\Vert
\nonumber \\
& & \times \left\Vert \tilde{U}_{D}(t) \prod_{j^\prime<k^\prime} f_{j^\prime k^\prime} (\bp_{\jp\kp} - \mu\sjpkp \bv\sjpkp) (1 + \vert  \tbx\sk - \tbx\sj \vert^2)^{-2} \right\Vert,
\label{Art1eq 2.8}
\end{eqnarray*}
\begin{eqnarray*}
I_2 &=& \left\Vert V\sjk^{vs} (\tbx\sk - \tbx\sj) g\sjk (\bp\sjk - \mu\sjk \bv\sjk) F(|\tilde\bx\sk - \tilde\bx\sj - \bv\sjk t - \be_1 q\sjk Et^2/(2\mu\sjk)| \geq \lambda | v\sjk t | 5/8) \, e^{-itH\scero} \right\Vert \nonumber \\
& & \times \left\Vert F(|\tilde\bx\sk - \tilde\bx\sj | \geq \lambda | v\sjk t |/8) \, \tilde{U}_{D}(t) \prod_{j^\prime<k^\prime} f_{j^\prime k^\prime} (\bp_{\jp\kp} - \mu\sjpkp \bv\sjpkp) (1 + \vert  \tbx\sk - \tbx\sj \vert^2)^{-2} \right\Vert, \label{Art1eq 2.9}
\end{eqnarray*}
\begin{eqnarray*}
I_3 &=& \left\Vert V\sjk^{vs} (\tbx\sk - \tbx\sj) g\sjk (\bp\sjk - \mu\sjk \bv\sjk) F(|\tilde\bx\sk - \tilde\bx\sj - \bv\sjk t - \be_1 q\sjk Et^2/(2\mu\sjk) | < \lambda | v\sjk t | 5/8) \right\Vert \nonumber \\
& & \times \left\Vert e^{-itH\scero} \, \tilde{U}_{D}(t) \prod_{j^\prime<k^\prime} f_{j^\prime k^\prime} (\bp_{\jp\kp} - \mu\sjpkp \bv\sjpkp) (1 + \vert  \tbx\sk - \tbx\sj \vert^2)^{-2} \right\Vert. \label{Art1eq 2.10}
\end{eqnarray*}

We give the proof for the pair $(1,2),$ the other cases are similar, using Jacobi coordinates based in the pair $(j,k).$ Let us set $\bx = \tbx_2 - \tbx_1, \bp = -i \nabla_\bsx$  we obtain as in \eqref{Art1eq 1.15} that
$$H\scero = \left[ (2\nu_1)^{-1} \bp^2 + q_{12}  \bE \cdot \bx \right] \otimes I + I \otimes \hat {H}\scero,$$ where $I \otimes \hat {H}\scero$ conmutes with $\bx$, by virtue of $\hat {H}\scero\textrm{'s}$ independence from $\bx.$ Note that $\nu_1 = \mu_{12}.$ Let us write $v = v_{12}=\vert \bv \vert$. Therefore, thanks to commutativity
\begin{eqnarray*}
e^{-it H\scero} &=& e^{-it \left[ (2\nu_1)^{-1} \bsp^2 + q_{12}  \bsE \cdot \bsx \right] \otimes I - it I \otimes \hat {H}\scero} = e^{-it \left[ (2\nu_1)^{-1} \bsp^2 + q_{12}  \bsE \cdot \bsx \right]} \otimes e^{ - it \hat {H}\scero}.
\end{eqnarray*}

We observe that the second factor in the tensorial product above conmutes with any operator depending on $\bx$ and $\bp$. It is also unitary, thus it disappears from the following norm estimations. We define $\mathcal{M}' = \{\bx \in \Rn \, \big| \, | \bx - \bv t | \geq \lambda | v t | 5/8  \}$ and $\mathcal{M} = \{\bx \in \Rn \, \big| \, |\bx | < \lambda | v t |/8  \}.$ We proceed as in Weder \cite{Weder1996MISCIAEF} using \eqref{Art1eq 2.1}-\eqref{Art1eq 2.4}.
\begin{eqnarray}
I_1
& \leq & C \left\Vert F(|\bx - \bv t - \be_1 q_{12}Et^2/(2\mu_{12}) | \geq \lambda | v t | 5/8) \, e^{-itH\scero} g_{12} (\bp - \mu_{12} \bv)  F(|\bx | < \lambda | v t |/8) \right\Vert \nonumber \\
& = & C \left\Vert F(|\bx - \bv t - \be_1 q_{12}Et^2/(2\mu_{12}) | \geq \lambda | v t | 5/8) \, e^{-i\bsp \cdot \bse_1 q_{12}Et^2/(2\mu_{12})} \right. \nonumber \\
&   & \ \ \left. \times e^{-it\bsp^2/(2\mu_{12})} g_{12} (\bp - \mu_{12} \bv)  F(|\bx | < \lambda | v t |/8) \right\Vert \nonumber \\
& = & C \left\Vert F(| \bx - \bv t | \geq \lambda | v t | 5/8) \, e^{-it\bsp^2/(2\mu_{12})} g_{12} (\bp- \mu_{12} \bv) F(|\bx | < \lambda | v t |/8) \right\Vert \nonumber \\
& = & C \Vert F(\bx \in \mathcal{M}') \, e^{-it\bsp^2/(2\mu_{12})} g_{12} (\bp- \mu_{12} \bv) F(\bx \in \mathcal{M}) \Vert \nonumber \\
&\leq& C(1 + \lambda |vt|/4 + |t|)^{-3}
\leq C(1 + |vt|)^{-3}.
\label{Art1eq 2.6 05}
\end{eqnarray}
To justify \eqref{Art1eq 2.6 05}, we will prove that $r \geq \lambda |vt|/4$ in
Proposition \ref{Art0pr 2.1}, provided $v > 4 \eta/ \lambda$.  Let us take $x \in \mathcal{M}'$ and $y \in \mathcal{M}+\bv t,$ then $| \bx - \by | = | \left( \bx - \bv t \right) - \left( \by - \bv t \right) | \geq \lambda| v t | 5/8 - \lambda | v t |/8 = \lambda | v t |/2.$ Thus, $r \geq \lambda | v t |/2 - \eta |t| \geq \lambda | v t |/2 - \lambda | v t |/4.$

Application of Lemma \ref{Art2le 3.1}, equation \eqref{Art2eq 3.3}, yields for an $\epsilon > 0,$
\beq\label{Art1eq 2.12}
I_2 \leq C(1 + |vt|)^{-2 - \epsilon}.
\ene
\begin{eqnarray}\label{Art1eq 2.13}
\textrm{Then, by \eqref{Art1eq 2.13 b},} \quad I_3
&\leq& C \Vert V^{vs}_{12} (\bx)g(\bp) F(\vert \bx - \bv t -\be_1 q_{12}Et^2/(2\mu_{12}) \vert < \lambda \vert vt \vert 5/8) \Vert \nonumber \\
&\leq& C \Vert V^{vs}_{12} (\bx)g(\bp) F(\vert \bx \vert \geq \vert vt \vert (\sqrt{\delta_1}-5\lambda/8)) \Vert \nonumber \\
&:=& h_{12}(\vert vt \vert),
\end{eqnarray}
where, by \eqref{Art1eq 2.5},
$h_{12}(\tau) \in L^1((0,\infty))$, provided $\lambda < 8 \sqrt{\delta_1}/5$.

Inequalities \eqref{Art1eq 2.8 -05},
\eqref{Art1eq 2.6 05}, \eqref{Art1eq 2.12} and \eqref{Art1eq 2.13} prove the Lemma.
\end{dem}


\begin{lemma}\label{Art2le 2.2}
Given $V^{E,\,s}_{jk} \in \Vscr_{E,\,s},$ where $1 \leq j < k \leq N,$ $\alpha$ as in Definition \ref{Art2df 1.2}, $\tilde{U}_{D}(t)$ be given as in \eqref{Art1eq 1.24}. Then for all functions $f_{j^\prime k^\prime} \in C_0^\infty (B_{\mu\sjpkp \eta\sjpkp})$ with $1 \leq j^\prime < k^\prime \leq N,$ there is a constant $0<c$ such that for every $\bv\sjk \in \R^n$ with $v\sjk > c,$ the following estimate is true for all $0 < \epsilon_1 < 1:$
\begin{eqnarray}\label{Art2eq 2.11}
\int\limits^\infty_{- \infty} dt \Bigg\Vert \left( V\sjk^{E,\,s} (\tbx\sk - \tbx\sj) - V\sjk^{E,\,s} (\bv\sjk t + \be_1 q\sjk E t^2/(2\mu\sjk)) \right)  & & \nonumber \\
\left.  \times
e^{-itH\scero} \tilde{U}_{D}(t) \prod_{j^\prime < k^\prime} f_{j^\prime k^\prime} (\bp_{\jp\kp} - \mu\sjpkp \bv\sjpkp) (1 + \vert  \tbx\sk - \tbx\sj \vert^2)^{-2} \right\Vert
& = &
  \begin{cases}
    O(v\sjk^{-\alpha}), & \hbox{\rm{if }} \alpha < 1, \cr
    O(v\sjk^{-1+\epsilon_1}), & \hbox{\rm{if }} \alpha = 1.
  \end{cases} \phantom{abcde}
\end{eqnarray}
\end{lemma}

\begin{dem} The proof is quite similar to that of Lemma 2.2 in Adachi and Maehara \cite{AdachiMaehara2007}.
To simplify the notation let us assume, in this Lemma, that $q_{12} \neq 0$ and consider the pair (1,2), i.e. $\bx = \tbx_2 - \tbx_1, \bp = -i \nabla_\bsx.$ Let us take $g_{12} \in C_0^\infty (B_{\mu_{12}\eta})$ that satisfies $g_{12} \equiv 1$ on the support of $f_{12}$.

We simplify as follows, noting that $V_{12}^{E,\,s}$ is bounded:
\begin{eqnarray}\label{DemArt2eq 2.11}
I &=& \left\Vert \left( V_{12}^{E,\,s} (\bx) - V_{12}^{E,\,s} (\bv t + \be_1 q_{12} E t^2/(2\mu_{12})) \right) \, e^{-itH\scero} \tilde{U}_{D}(t)
\prod_{j^\prime<k^\prime} f_{j^\prime k^\prime} (\bp_{\jp\kp} - \mu\sjpkp \bv\sjpkp) (1 + \bx^2)^{-2} \right\Vert \nonumber \\
&\leq&
C (I_1 + I_2 + I_3), \nonumber
\end{eqnarray} where, for $0 < \tilde{\alpha} < 1,$
\begin{eqnarray}
I_1 &=& \left\Vert F(|\bx - \bv t - \be_1 q_{12} Et^2/(2\mu_{12}) | \geq 3 | v^{\tilde{\alpha}} t |) \, e^{-itH\scero} g_{12} (\bp - \mu_{12} \bv)  F(|\bx| < | v^{\tilde{\alpha}} t |) \right\Vert
\nonumber \\
&=& \left\Vert F(| \bx - \bv t | \geq 3 | v^{\tilde{\alpha}} t |) \, e^{-it\bsp^2/(2\mu_{12})} g_{12} (\bp- \mu_{12} \bv) F(|\bx | < | v^{\tilde{\alpha}} t |) \right\Vert, \nonumber
\\
I_2 &=& \left\Vert F(|\bx| \geq | v^{\tilde{\alpha}} t |)  \tilde{U}_{D}(t) \prod_{j^\prime<k^\prime} f_{j^\prime k^\prime} (\bp_{\jp\kp} - \mu\sjpkp \bv\sjpkp) (1 + \bx^2)^{-2} \right\Vert, \nonumber
\\
I_3 &=& \left\Vert \left( V_{12}^{E,\,s} (\bx) - V_{12}^{E,\,s} (\bv t + \be_1 q_{12} E t^2/(2\mu_{12})) \right) 
F(|\bx - \bv t - \be_1 q_{12} Et^2/(2\mu_{12}) | < 3| v^{\tilde{\alpha}} t |) \right\Vert \nonumber \\
&=& \left\Vert \left( V_{12}^{E,\,s} (\bx + \bv t + \be_1 q_{12} Et^2/(2\mu_{12}) ) - V_{12}^{E,\,s} (\bv t + \be_1 q_{12} E t^2/(2\mu_{12})) \right)
F(|\bx | < 3| v^{\tilde{\alpha}} t |) \right\Vert. \nonumber
\end{eqnarray}

$I_1$ and $I_2$ are estimated as in the proof of Lemma \ref{Art2le 2.1}, by Proposition \ref{Art0pr 2.1} and equation \ref{Art2eq 3.3}, respectively:
$$
\int_{-\infty}^{\infty} \left(I_1 + I_2 \right) \, dt = O(v^{-\tilde{\alpha}}).
$$
By lemma 2.2 of Adachi and Maehara \cite{AdachiMaehara2007} (see also page 042101-5, equation 2.10 of \cite{AdachiMaehara2007}), we get, for all $0 < \tilde{\alpha} < 1$ and $v$ sufficiently large that
$$
\int_{-\infty}^{\infty} I_3 \, dt =
  \begin{cases}
    O(v^{\tilde{\alpha}-2\alpha}), & \textrm{if } \alpha < 1, \cr
    O(v^{\tilde{\alpha}-2} |\ln v| ), & \textrm{if } \alpha = 1.
  \end{cases}
$$

We finish the proof by setting
$
\tilde{\alpha} =
  \begin{cases}
    \alpha, & \textrm{if } \alpha < 1, \cr
    1 - \epsilon_1, & \textrm{if } \alpha = 1, \, 0 < \epsilon_1 < 1.
  \end{cases}
$
\end{dem}

\begin{lemma}\label{Art2le 3.4}
Let $V\sjk^{l}$ and $\tilde{U}_{D}(t)$ be given as in \eqref{Art1eq 1.23 -25} and \eqref{Art1eq 1.24}, respectively. Let $\gamma_1, \epsilon_0$ be as in Definition \ref{Art2df 1.1}, $\gamma_D, \mu$ be as in Definition \ref{Art2df 1.2}, $\theta\sjk$ as in \eqref{Art2eq 3.7 -10}. Let us define two constants ${\sigma\sjk}$ and $\tilde\sigma\sjk;$ if $q\sjk \neq 0$ and $V\sjk^{l} \neq 0,$ then
${\sigma\sjk} = \frac{\tilde\sigma\sjk}{2-{\tilde\sigma\sjk}}$
and $0 < {\tilde\sigma\sjk} < 2 - \max \{\frac{1+\theta\sjk}{\gamma_D + \mu}, \, \frac{2}{\gamma_D + 2\mu}, \, 1
\},$ else, if $q\sjk = 0$ or $V\sjk^{l} = 0,$ then ${\sigma\sjk}:={\tilde\sigma\sjk}:=1.$ Then for all functions $f_{j^\prime k^\prime} \in C_0^\infty (B_{\mu\sjpkp \eta\sjpkp})$ with $1 \leq j^\prime < k^\prime \leq N,$ and for all integers $1 \leq j < k \leq N,$ there is a constant $v_0>0$ such that for every $\bv\sjk \in \R^n$ with $v\sjk > v_0^{1/\sigma\sjk},$
we have the following estimate:
\begin{eqnarray}\label{Art2eq 3.7}
\int\limits^\infty_{- \infty} dt \, \Bigg\Vert \left( V\sjk^{l} (\tbx\sk - \tbx\sj) - V\sjk^{l} ( t \, \bp\sjk/\mu\sjk - \be_1 q\sjk E t^2/(2\mu\sjk)) \right)
 \nonumber \\
\left. \times
\, e^{-itH\scero} \tilde{U}_{D}(t)  \prod_{j^\prime<k^\prime} f_{j^\prime k^\prime} (\bp_{\jp\kp} - \mu\sjpkp \bv\sjpkp) (1 + \vert  \tbx\sk - \tbx\sj \vert^2)^{-2} \right\Vert
& \leq &
    O(v\sjk^{-{\sigma\sjk}}).
\end{eqnarray}
\end{lemma}
\begin{dem}
The proof in the case where $q\sjk = 0$ is quite similar to that of Lemma 3.3 in Enss and Weder \cite{EnssWeder1995}, and the proof in the case where $q\sjk \neq 0$ is quite similar to that of Lemma 3.4 in Adachi and Maehara \cite{AdachiMaehara2007}.
In this Lemma, let us
denote $\bx = \tbx\sk - \tbx\sj, \bp = -i \nabla_\bsx.$
From \eqref{Art1eq 2.13 b3}, a constant is defined as follows
$$
c :=
\begin{cases}
c_1^{\tilde\sigma\sjk} c_2^{1-\tilde\sigma\sjk}, & \textrm{if } q\sjk \neq 0, \\
1/2, & \textrm{if } q\sjk = 0.
\end{cases}
$$

Let's split the long-range potential $V\sjk^{l}$ into two parts with controllable decay properties. Let $\chi \in C^\infty(\Rn)$ satisfy, $0 \leq \chi \leq 1, \chi (\bu) = 1,$ for $\vert \bu \vert \geq c$ and $\chi (\bu) = 0,$ for $\vert \bu \vert \leq c/2;$ and $V_{jk,v\sjk t}^{l}(\bu)=V\sjk^{l}(\bu) \chi (\bu /(v\sjk^{{\tilde\sigma\sjk}} \vert t \vert^{2-{\tilde\sigma\sjk}})).$ In consequence, $\textrm{supp} \left(V_{jk,v\sjk t}^{l}-V\sjk^{l}\right) \subset B_{c v\sjk^{{\tilde\sigma\sjk}} \vert t \vert^{2-{\tilde\sigma\sjk}}},$ and $\Vert V_{jk,v\sjk t}^{l}-V\sjk^{l} \Vert \leq \Vert V\sjk^{l} \Vert.$

Choosing again $g \in C_0^\infty (B_{\mu\sjk\eta\sjk})$ such that $g \equiv 1$ on the support of $f\sjk$, it follows that
\begin{eqnarray}
& & \left\Vert \left( V\sjk^{l} (\bx) - V\sjk^{l} (t \, \bp/\mu\sjk - \be_1 q\sjk E t^2/(2\mu\sjk)) \right) e^{-itH\scero} \tilde{U}_{D}(t) \prod_{j^\prime<k^\prime} f_{j^\prime k^\prime} (\bp_{\jp\kp} - \mu\sjpkp \bv\sjpkp) (1 + \bx^2)^{-2} \right\Vert\nonumber \\
&
\leq
&
I_1 + I_2 + I_3,
\end{eqnarray}
where
\begin{eqnarray}
I_1
& = & \left\Vert \left( V_{jk,v\sjk t}^{l} (\bx) - V\sjk^{l} (t \, \bp/\mu\sjk - \be_1 q\sjk E t^2/(2\mu\sjk)) \right) e^{-itH\scero} g(\bp - \mu\sjk \bv\sjk) \phantom{\prod_{j^\prime<k^\prime}} \right.
\nonumber \\
&   & \times
\left. \tilde{U}_{D}(t) \prod_{j^\prime<k^\prime} f_{j^\prime k^\prime} (\bp_{\jp\kp} - \mu\sjpkp \bv\sjpkp) (1 + \bx^2)^{-2} \right\Vert, \label{Art1eq 2.10 lr}
\end{eqnarray}
\begin{eqnarray}
I_2
& = & \left\Vert \left( V\sjk^{l} - V_{jk,v\sjk t}^{l} \right)(\bx) \, e^{-itH\scero} g(\bp - \mu\sjk \bv\sjk) F(| \bx | < v\sjk^{{\sigma\sjk}} | t |/8) \right\Vert
\nonumber \\
&   & \times
\left\Vert \tilde{U}_{D}(t) \prod_{j^\prime<k^\prime} f_{j^\prime k^\prime} (\bp_{\jp\kp} - \mu\sjpkp \bv\sjpkp) (1 + \bx^2)^{-2} \right\Vert, \label{Art1eq 2.8 lr}
\end{eqnarray}
\begin{eqnarray}
I_3
& = & \left\Vert \left( V\sjk^{l} - V_{jk,v\sjk t}^{l} \right)(\bx) \, e^{-itH\scero} g(\bp - \mu\sjk \bv\sjk) \right\Vert \nonumber \\
&   & \times \left\Vert F(| \bx | \geq v\sjk^{{\sigma\sjk}} | t |/8) \tilde{U}_{D}(t) \prod_{j^\prime<k^\prime} f_{j^\prime k^\prime} (\bp_{\jp\kp} - \mu\sjpkp \bv\sjpkp) (1 + \bx^2)^{-2} \right\Vert. \label{Art1eq 2.9 lr}
\end{eqnarray}

If $q\sjk \neq 0,$ for $\bq \in B_{\mu\sjk\eta\sjk}, \,
v\scero \geq 4 \eta\sjk /\sqrt{1-\delta\sjk},$ by \eqref{Art1eq 2.13 b3}, we have
\begin{eqnarray}
\vert t \, \bq/\mu\sjk + \bv\sjk t + \be_1 q\sjk E t^2/(2\mu\sjk) \vert &\geq& c v\sjk^{{\tilde\sigma\sjk}} \vert t \vert^{2-{\tilde\sigma\sjk}}.
\label{Art2eq 3.9 05}
\end{eqnarray}

If  $q\sjk \neq 0$ with $\bp$ in the support of $g,$ we note, by \eqref{Art2eq 3.9 05}, that $V_{jk,v\sjk t}^{l} (t \, \bp/\mu\sjk + \bv\sjk t + \be_1 q\sjk E t^2/(2\mu\sjk)) = $ $ V\sjk^{l} (t \, \bp/\mu\sjk + \bv\sjk t + \be_1 q\sjk E t^2/(2\mu\sjk)).$ If $q\sjk = 0,$ $\bp$ belongs to the support of $g ( \cdot - \mu\sjk \bv\sjk),$ and $v_0 > 2 \eta\sjk$ then $V_{jk,v\sjk t}^{l} (t \, \bp/\mu\sjk) = V\sjk^{l} (t \, \bp/\mu\sjk).$

As in Enss and Weder \cite{EnssWeder1995} and Adachi and Maehara \cite{AdachiMaehara2007}, by \eqref{Art1eq 2.1}-\eqref{Art0eq 3.14 05} and
the Baker-Campbell-Hausdorff formula \cite{Enss1979},
\begin{eqnarray}
I_1
&
\leq &\int_0^1 ds \left\Vert \bigg[ \left( \nabla V_{jk,v\sjk t}^{l} \right)(s\bx + t \, \bp/\mu\sjk + \bv\sjk t + \be_1 q\sjk E t^2/(2\mu\sjk)) \cdot \bx \right.
\nonumber \\
& & \ \ \left. \left.
+ \frac{it}{(2\mu\sjk)} \left( \Delta V_{jk,v\sjk t}^{l} \right)(s\bx + t \, \bp/\mu\sjk
 + \bv\sjk t + \be_1 q\sjk E t^2/(2\mu\sjk)) \right] g(\bp) \, e^{-i \mu\sjk \bsv\sjk \cdot \bsx}
\right. \nonumber \\
& & \ \ \left. \times
\tilde{U}_{D}(t) \prod_{j^\prime<k^\prime} f_{j^\prime k^\prime} (\bp_{\jp\kp} - \mu\sjpkp \bv\sjpkp) (1 + \bx^2)^{-2} \right\Vert. \label{Art0eq 3.9}
\end{eqnarray}

For $q\sjk=0,$ $\epsilon_0 < \gamma_1 - 3/2,$ having in consideration that in the support of $V_{jk,v\sjk t}^{l}$ we must have $\vert \bx \vert \geq (c/2) \vert v\sjk t \vert,$ \eqref{Art2eq 3.2b} and \eqref{Art0eq 3.9} imply that
\begin{eqnarray}
I_1
& \leq & C \left[ \vert v\sjk t \vert^{-\gamma_1}
\left( 1 + v\sjk^{-(2-\gamma_1/2)}|v\sjk t|^{\theta\sjk}  \right) + \vert v\sjk t \vert^{-1-2\epsilon_0}  \right] \ \leq \ C \vert v\sjk t \vert^{-1-2\epsilon_0}. \nonumber \end{eqnarray}
Then, by the fact that $I_1$ is uniformly bounded, for all $t \in \R$ and all $v\sjk,$
$
\int
dt \, I_1
=
O(v\sjk^{-1}).
$

We consider now the case when $q\sjk\neq0.$ Recall that in the support of $V_{jk,v\sjk t}^{l}$ we must have $\vert \bx \vert \geq (c/2) v\sjk^{{\tilde\sigma\sjk}} \vert t \vert^{2-{\tilde\sigma\sjk}}$ for $0 < {\tilde\sigma\sjk} < 1.$
For
$0<b,$
\eqref{Art2eq 3.2b}
and \eqref{Art0eq 3.9}
imply:
$$
I_1 \leq I_{11} + I_{12},
\textrm { where, } I_{12} \leq C \Vert V\sjk^{l} \Vert,
$$
\begin{eqnarray}
I_{11}
&\leq& C \left( \left( v\sjk^{-{\tilde\sigma\sjk} (\gamma_D+\mu)} \vert t \vert^{-(2-{\tilde\sigma\sjk})(\gamma_D + \mu)}
+  v\sjk^{-{\tilde\sigma\sjk}(\gamma_D+1)} \vert t \vert^{-(2-{\tilde\sigma\sjk})(\gamma_D+1)}
\right)
\right.
 \nonumber \\
&    & \times \left.
\left(1 + v\sjk^{-(2-\gamma_1/2)} |v\sjk t|^{\theta\sjk}\right)
F(|t| > v\sjk^{-b}
) + \Vert V\sjk^{l} \Vert  \, F(|t| \leq v\sjk^{-b})\right) \textrm {  and }
\nonumber
\\
I_{12}
&\leq&
C \left( v\sjk^{-{\tilde\sigma\sjk}(\gamma_D+2\mu)} \vert t \vert^{-(2-{\tilde\sigma\sjk})(\gamma_D+2\mu)+1} +  v\sjk^{-{\tilde\sigma\sjk}(\gamma_D+\mu+1)} \vert t \vert^{-(2-{\tilde\sigma\sjk})(\gamma_D+\mu+1)+1} \right.
\nonumber \\
&    &
\left.
\
+ v\sjk^{-{\tilde\sigma\sjk}(\gamma_D+2)} \vert t \vert^{-(2-{\tilde\sigma\sjk})(\gamma_D+2)+1} \right).
\nonumber
\end{eqnarray}

By a straightforward calculation, provided ${\tilde\sigma\sjk} < 2 - (1+\theta\sjk)/(\gamma_D + \mu),$
$$
\int
dt \, I_{11}
=
O(v\sjk^{-b}),
$$
having taken
$$b = \frac{{\tilde\sigma\sjk}}{2-{\tilde\sigma\sjk}} = \min \left\{\frac{{\tilde\sigma\sjk}}{2-{\tilde\sigma\sjk}}, \, \frac{{\tilde\sigma\sjk} + (2-\gamma_1/2 - \theta\sjk)/ (\gamma_D+1)}{2-{\tilde\sigma\sjk} - \theta\sjk/(\gamma_D+1)}, \,  \frac{{\tilde\sigma\sjk} + (2-\gamma_1/2 - \theta\sjk)/ (\gamma_D+\mu)}{2-{\tilde\sigma\sjk}- \theta\sjk/(\gamma_D+\mu)} \right\}.$$

Using Adachi and Maehara's computations \cite{AdachiMaehara2007} of the last three terms of the integral of $I_3$ in the proof of their Lemma 3.4, assuming that ${\tilde\sigma\sjk} < 2 - 2/(\gamma_D + 2\mu),$ we obtain:
\begin{eqnarray}
\int_{-\infty}^{+\infty} dt \, I_{12}
&=&
O(v\sjk^{-{\tilde\sigma\sjk}/\left[(2-{\tilde\sigma\sjk})-1/(\gamma_D+2)\right]}). \nonumber
\end{eqnarray}

Thus we have, in general:
\begin{eqnarray}
\int_{-\infty}^{+\infty} dt \, I_1
& = & O(v\sjk^{-{\sigma\sjk}}). \label{Art0eq 3.12 30}
\end{eqnarray}

If $\vert \bx \vert \leq (5/8) \, v\sjk^{\sigma\sjk} \vert t \vert$
and $v\sjk^{{\sigma\sjk} - 1} \leq (2/5)\sqrt{1-\delta\sjk},$
for $q\sjk \neq 0$, we obtain as in \eqref{Art2eq 3.9 05}
\begin{eqnarray}
\vert \bx + \bv\sjk t + \be_1 q\sjk E t^2/(2\mu\sjk) \vert &\geq& c v\sjk^{{\tilde\sigma\sjk}} \vert t \vert^{2-{\tilde\sigma\sjk}}, \label{Art2eq 3.9 10}
\end{eqnarray}
by equations \eqref{Art1eq 2.1}-\eqref{Art1eq 2.4} and \eqref{Art2eq 3.9 10} we can invoke Proposition 2.10 from Enss \cite{Enss1983} in \eqref{Art1eq 2.8 lr} to estimate $I_2$ with $v_0 > 4 \eta\sjk$,
\begin{eqnarray}
I_2
& \leq &
C \left\Vert \left( V\sjk^{l} - V_{jk,v\sjk t}^{l} \right)(\bx + \bv\sjk t + \be_1 q\sjk Et^2/(2\mu\sjk)) \, e^{-it\bsp^2/(2\mu\sjk)} g(\bp) F(| \bx | < v\sjk^{{\sigma\sjk}} | t |/8) \right\Vert \nonumber \\
& \leq &
C \left\Vert F\left(| \bx - \bv\sjk t  | \geq
    \begin{cases}
    5 v\sjk^{{\sigma\sjk}} | t |/8, & \textrm {if } q\sjk \neq 0, \\
    v\sjk | t |/2, & \textrm {if } q\sjk = 0, \\
    \end{cases} \right)
    \, e^{-it\bsp^2/(2\mu\sjk)} g(\bp - \mu\sjk \bv\sjk) F(| \bx | < v\sjk^{{\sigma\sjk}} | t |/8) \right\Vert
\nonumber \\
& \leq & C (1 + v\sjk^{\sigma\sjk} \vert t \vert)^{-2}. \label{Art1eq 2.11 lr a}
\end{eqnarray}

Again, by Lemma \ref{Art2le 3.1}, equation \eqref{Art2eq 3.3}, we estimate $I_3,$
\beq
I_3 \leq C (1 + v\sjk^{\sigma\sjk} \vert t \vert)^{-2}. \label{Art1eq 2.11 lr b}
\ene

By
\eqref{Art0eq 3.12 30}, \eqref{Art1eq 2.11 lr a} and \eqref{Art1eq 2.11 lr b} we finish the proof.
\end{dem}

Let us denote,
\beq\label{Art1eq 3.20 05} I_{G,v,a,b}= \exp\left(- i \sum\limits_{j < k}^E
\int^b_a ds\; V^{E,\,s}_{jk}(\bv\sjk s + \be_1 q\sjk E s^2/(2\mu\sjk))\right) \textrm{ and } I_{G,v}=I_{G,v,-\infty,\infty}. \ene
Observe that $\tilde{U}_{G,v} (t)=I_{G,v,0,t}.$

\begin{prop}\label{DollardGrafWaveOperators}
The wave operators $\Omega_\pm^{D,G,v}$ and $\Omega_\pm^{G,v}$ exist and, moreover,
\beq\label{Art1eq 3.24 -10}
\Omega_\pm^{D,G,v} = W_\pm^{D} I_{G,v,0,\pm \infty}, \quad \Omega_\pm^{G,v} = W_\pm^{} I_{G,v,0,\pm \infty}.
\ene
\end{prop}
\begin{dem}
We give only the proof for $\Omega_\pm^{D,G,v},$ the other is similar. Note that:
$$
\textrm{s}-\lim_{t \to \pm \infty} e^{itH} U^{D,G,v} (t)
=  \textrm{s}-\lim_{t \to \pm \infty} e^{itH} e^{-itH\scero} \tilde{U}_{D} (t) I_{G,v,0,t} \\
=  W_\pm^{D} \ \textrm{s}-\lim_{t \to \pm \infty} I_{G,v,0,t}.
$$
Furthermore, for any $\Phi \in L^2$ we have that:
$$
\Vert \left( I_{G,v,0,t} - I_{G,v,0,\pm \infty} \right) \Phi \Vert^2
 =  \int_{\Rn} \vert I_{G,v,0,t} - I_{G,v,0,\pm \infty} \vert^2 \vert \Phi \vert^2  \longrightarrow_{t \to \pm \infty} 0,
$$ by the Lebesgue dominated convergence theorem, taking into account that the integrand is dominated by $4\vert \Phi \vert^2$ for all $t$. This proves the proposition.
\end{dem}

Now we focus in the wave operator estimates. We use Jacobi Coordinates based on the pair (1,2), where $v = \vert \bv \vert = \vert \bv_2 - \bv_1 \vert$ and $v\sjk = O (v^2),$ for $(j,k) \neq (1,2).$ Lemma \ref{Art2le 2.3}, below, is a N-body generalization of Lemma 3.5 in Adachi and Maehara \cite{AdachiMaehara2007}.
See also Lemma 4.6 of Adachi, Kamada, Kazuno and Toratani \cite{AdachiEtAl2011}, for a generalization of Lemma 3.5 of Adachi and Maehara \cite{AdachiMaehara2007} to the case where the external electric field is asymptotically zero in time, in the two-body case.


\begin{lemma}\label{Art2le 2.3}
Let
$\alpha
$ be as in Definition \ref{Art2df 1.2}, where, without loss of generality, $\alpha=1$ if $q\sjk = 0$ for all $1 \leq j < k \leq N.$
For all $1 \leq j < k \leq N,$ let $0 < \sigma\sjk \leq 1$ be as in Lemma \ref{Art2le 3.4}.
Let us take $V^{VS} \in \Vscr_{VSR},V^S \in \Vscr_{SR},V^L \in \Vscr_{LR}.$ Then, for all $\Phi_{\bsv}$ as in \eqref{Art1eq 3.3} with a fixed normalized $\hat{\phi}_{3}$, where, with $\delta\sjk$ being defined as in \eqref{Art1eq 2.13 c}, the relative velocities satisfy $\vert \hat{\bv}\sjk \cdot \hat{\bE} \vert \leq \delta\sjk$ for all integers $1 \leq j < k \leq N$ with $q_{j,k} \neq 0,$ and $v\sjk > v_0^{1/\sigma\sjk}
$ for some $v_0 > 0$ and all integers $1 \leq j < k \leq N:$
\begin{eqnarray}
\sup_{t \in \R} \left\Vert (\Omega_\pm^{D,G,v} - e^{itH}U^{D,G,v}(t))\Phi_{\bsv} \right\Vert
&=&
O (v^{-\min \{\alpha, \, \sigma\sjk  \, | \, 1 \leq j < k \leq N \}}).
\phantom{abcde} \label{Art1eq 3.20}
\end{eqnarray}
%
%
%
In the short-range case, where $V^{l}\sjk = 0$ (see \eqref{Art1eq 1.23 -25}) for all $1 \leq j < k \leq N,$ we obtain the following result
\begin{eqnarray}
\sup_{t \in \R} \left\Vert (\Omega_\pm^{G,v} - e^{itH}U^{G,v}(t))\Phi_{\bsv} \right\Vert
&=&
        \begin{cases}
            O (v^{-\alpha}), & \hbox{\rm{if }} \alpha < 1 \hbox{\rm{ and }} \sum\limits_{j < k} \vert q\sjk \vert > 0, \\
            O (v^{-(1-\epsilon_1)}), & \hbox{\rm{if }} \alpha = 1 \hbox{\rm{ and }} \sum\limits_{j < k} \vert q\sjk \vert > 0, \phantom{espacio}\\
            O (v^{-1}), & \hbox{\rm{if }} \sum\limits_{j < k} \vert q\sjk \vert = 0,
         \end{cases}
\phantom{abcde} \label{Art1eq 3.20b}
\end{eqnarray} for all $0 < \epsilon_1 < 1.$
\end{lemma}

\begin{dem}
We give the proof for $\Omega_+^{D,G,v}$. By Duhamel's formula, \eqref{Art1eq 2.1} and \eqref{Art1eq 2.3}:
\begin{eqnarray*}
\Omega_+^{D,G,v} - e^{itH}U^{D,G,v}(t) & = & \lim_{t' \to + \infty} e^{it'H} U^{D,G,v}(t') - e^{itH}U^{D,G,v}(t)
=
\lim_{t' \to + \infty} \int_t^{t'} ds \; \frac{d}{ds} \left( e^{isH} U^{D,G,v}(s) \right)
\\
\phantom{\Omega_+^{D,G,v} - e^{itH}U^{D,G,v}(t)}
& = & i \int_t^{\infty} ds \; e^{isH} \left( \sum_{j<k} \left[ V_{jk}^{vs} (\tbx\sk - \tbx\sj)
+  V_{jk}^{l} (\tbx\sk - \tbx\sj) \right.
\right. \\
&   & \left.
- V_{jk}^{l}(s \bp\sjk/\mu\sjk - \be_1 q\sjk E s^2/(2\mu\sjk))
\right] \\
&   & \left.  + \sum_{j<k}^E \left[ V_{jk}^{E,\,s} (\tbx\sk - \tbx\sj) -
 V^{E,\,s}_{jk}(\bv\sjk s + \be_1 q\sjk E s^2/(2\mu\sjk)) \right] \right) U^{D,G,v}(s). \\
\end{eqnarray*}
Using that the Graf-type modifier $\tilde{U}_{G,v} (t)$ \eqref{Art1eq 1.23 -05} conmutes with any operator, and Lemmata \ref{Art2le 2.1}, \ref{Art2le 2.2}, \ref{Art2le 3.4}, it follows for any $0 < \epsilon_1 < 1, \, t \in \R$:
\begin{eqnarray*}
\left\Vert \left( e^{-itH}\Omega_+^{D,G,v} - U^{D,G,v}(t) \right) \Phi_{\bsv} \right\Vert
&\leq& C \sum\limits_{j < k} \int_{-\infty}^{\infty} \; \Big\Vert V^{vs}_{jk}(\tbx\sk-\tbx\sj)  \\
& & \ \   \times  e^{-isH\scero} \tilde{U}_{D}(s)  \prod_{j^\prime<k^\prime} f_{j^\prime k^\prime} (\bp_{\jp\kp} - \mu\sjpkp \bv\sjpkp)
(1 + \vert  \tbx\sk - \tbx\sj \vert^2)^{-2} \Big\Vert \; ds
\\
& & +  C \sum\limits_{j < k} \int_{-\infty}^{\infty} \Big\Vert \left( V_{jk}^{l} (\tbx\sk - \tbx\sj)
- V_{jk}^{l}(s \bp\sjk/\mu\sjk  - \be_1 q\sjk E s^2/(2\mu\sjk))\right)  \\
& & \ \  \times  e^{-isH\scero} \tilde{U}_{D}(s)  \prod_{j^\prime<k^\prime} f_{j^\prime k^\prime} (\bp_{\jp\kp} - \mu\sjpkp \bv\sjpkp)
(1 + \vert  \tbx\sk - \tbx\sj \vert^2)^{-2} \Big\Vert \; ds
\\
& & +  C \sum\limits_{j < k}^E \int_{-\infty}^{\infty} \Big\Vert \left(  V^{s}_{jk}(\tbx\sk-\tbx\sj) - V^{s}_{jk}(\bv\sjk s + \be_1 q\sjk E s^2/(2\mu\sjk))\right) \\
& & \ \  \times e^{-isH\scero} \tilde{U}_{D}(s)  \prod_{j^\prime<k^\prime} f_{j^\prime k^\prime} (\bp_{\jp\kp} - \mu\sjpkp \bv\sjpkp) (1 + \vert  \tbx\sk - \tbx\sj \vert^2)^{-2} \Big\Vert \; ds
\\
\phantom{\left\Vert \left( e^{-itH}\Omega_+^{D,G,v} - U^{D,G,v}(t) \right) \Phi_{\bsv} \right\Vert}
&\leq& \sum\limits_{j < k} \left( O(v\sjk^{-1}) + O(v\sjk^{-\sigma\sjk}) \right) + \sum\limits_{j < k}^E
  \begin{cases}
    O(v\sjk^{-\alpha}), & \textrm{if } \alpha < 1, \cr
    O(v\sjk^{-(1-\epsilon_1)} ), & \textrm{if } \alpha = 1.
  \end{cases} \\
\end{eqnarray*} The proof is finished by the use of the following arguments: $\alpha < 1$ implies for $v \geq 1$ that $ v^{-1} \leq v^{-\alpha},$
$0 < \sigma\sjk < 1$ implies, for $\epsilon_1$ sufficiently small that $O (v\sjk^{-(1-\epsilon_1)}) \leq O(v\sjk^{-\sigma\sjk}),$ and noting that $v_{12}=v$ and $v\sjk$ is $O(v^2)$ for $j<k=3, \ldots, N.$
\end{dem}

Lemma \ref{Art2le 3.4} above defines two sets of exponents $\sigma\sjk$ and $\tilde\sigma\sjk.$ Theorem \ref{Art2tr 2.1} below needs $\sigma\sjk > 1/2.$ For this purpose we have to ask, for all $1 \leq j < k \leq N$ with $q\sjk \neq 0$ and $V\sjk^{l} \neq 0,$ that $\theta\sjk,$ being as in \eqref{Art2eq 3.7 -10}, must hold:
\beq
2 - \max \{\frac{1+\theta\sjk}{\gamma_D + \mu}, \, \frac{2}{\gamma_D + 2\mu}, \, 1 \} > \frac{2}{3}
\Longleftrightarrow
\theta\sjk < \frac{4}{3} (\gamma_D + \mu) - 1.
\label{Art2tr 2.1 -80}
\ene
In particular, inequality \eqref{Art2tr 2.1 -80} is always true if $\theta\sjk \leq 1/3$ because $1/3 < 4 (\gamma_D + \mu)/3 - 1$
for all $\gamma_D$ and $\mu$ as in Definition \ref{Art2df 1.2}.
Inequality \eqref{Art2tr 2.1 -80} is always met in conditions $\zeta\sjk^{b}$ and $\zeta\sjk^{c},$ see \eqref{Art2eq 3.7 -10}, because in the former, $\theta\sjk$ can be taken arbitrarily small, and in the later, $\theta\sjk$ is zero. If there is a pair $(j,k)$ with $q\sjk \neq 0$ and the condition $\zeta\sjk^{a}$ is true, \eqref{Art2tr 2.1 -80} is equivalent to $\max \{3/2, \, 3 - 4 (\gamma_D + \mu)/3 \} < \gamma_1 < 2.$
If $\sum | q\sjk | = 0$ we just need $3/2 < \gamma_1 \leq 2.$ Theorem \ref{Art1te 1.2} is stated considering long-range potentials, in this case, $\zeta\sjk^{a}$ is true for some pair $(j,k)$ with $q\sjk \neq 0,$ if and only if, there are two pairs $(j^*,k^*)$ and $(j',k')$ such that $1 \leq j^*<k^* \leq N,$ $1 \leq j'<k' \leq N,$ $q_{j^*k^*} \neq 0$ and $q\sjpkp = 0.$ We can also use Theorem \ref{Art1te 1.2} with short-range potentials: the condition $3 - 4 (\gamma_D + \mu)/3 < \gamma_1$ is always true because, without loss of generality, we can take $\gamma_1=2$ in this situation.

\begin{theorem}\label{Art2tr 2.1} (Reconstruction Formula)
Let $\gamma_1$ be as in Definition \ref{Art2df 1.1}, $\alpha,\gamma_D,\mu$ be as in Definition \ref{Art2df 1.2}, where, without loss of generality, $\alpha=1$ if $q\sjk = 0$ for all $1 \leq j < k \leq N.$ If there exists two pairs $1 \leq j < k \leq N,$ $1 \leq j' < k' \leq N$ such that $q\sjk \neq 0,$ $q\sjpkp=0, \, V\sjpkp^{l} \neq 0,$ and either $\jp = j \textrm{ or } \jp = k \textrm{ or } \kp = j \textrm{ or } \kp = k \textrm{ or } \jp + j = 3,$ we additionally assume $\gamma_1 > 3 - 4 (\gamma_D + \mu)/3.$ For all $1 \leq j < k \leq N,$ let $0 < \sigma\sjk \leq 1$ be as in Lemma \ref{Art2le 3.4}. Let us take $V^{VS} \in \Vscr_{VSR},V^S \in \Vscr_{SR},V^L \in \Vscr_{LR},$ where $V_{12}^{vs}$ satisfies \eqref{Art1eq 2.5} for all $g \in C_0^\infty(\R^n)$, with $0 \leq \rho \leq 2\min \{\alpha, \, \sigma\sjk  \, | \, 1 \leq j < k \leq N \}-1.$ Let us set $\bp_l = \bp \cdot \be_l$ for any $l=1,\ldots,N.$
Then, for all $\Phi_{\bsv},\Psi_{\bsv}$ as in \eqref{Art1eq 3.3} with the same fixed normalized $\hat{\phi}_{3}$, with $\delta\sjk$ being defined as in \eqref{Art1eq 2.13 c}, the relative velocities satisfy $\vert \hat{\bv}\sjk \cdot \hat{\bE} \vert \leq \delta\sjk$ for all integers $1 \leq j < k \leq N$ with $q_{j,k} \neq 0,$ and $v\sjk > v_0^{1/\sigma\sjk}$ for some $v_0 > 0,$ as in Lemma \ref{Art2le 3.4},  and all integers $1 \leq j < k \leq N:$
\begin{eqnarray}\label{Art1eq 3.24}
v(i[S^{D}, \bp_l] \Phi_{\bsv}, \Psi_{\bsv})
& = & \int_{-\infty}^{\infty} d\tau \Big[ (V_{12}^{vs}(\bx + \tau \hat{\bv}) \bp_l \Phi_{12}, \Psi_{12})
-(V_{12}^{vs}(\bx + \tau \hat{\bv}) \Phi_{12}, \bp_l \Psi_{12}) \nonumber \\
&  & \ +i\left( \left( \frac{\partial V_{12}^{s}}{\partial x_l} \right)(\bx + \tau \hat{\bv}) \Phi_{12}, \Psi_{12} \right) +i\left( \left( \frac{\partial V_{12}^{l}}{\partial x_l} \right)(\bx + \tau \hat{\bv}) \Phi_{12}, \Psi_{12} \right)
\Big] \nonumber \\
& &
+
\begin{cases}
o(v^{-\rho_l}), & \hbox{\rm{if }} \gamma_2 - 1 \leq \rho \leq 2\min \{\alpha, \, \sigma\sjk  \, | \, 1 \leq j < k \leq N \} -1 < 1, \cr & \hbox{\rm{ for any $\rho_l$, }} 0 \leq \rho_l < \gamma_2 - 1, \\
o(v^{-\rho}), & \hbox{\rm{if }} 0 \leq \rho < \min \{ \gamma_2, 2\alpha, \, 2\sigma\sjk  \, | \, 1 \leq j < k \leq N \}-1, \\
O(v^{-\rho}), & \hbox{\rm{if }} \rho = 2\min \{\alpha, \, \sigma\sjk  \, | \, 1 \leq j < k \leq N \} -1 < \gamma_2 -1, \\
o(v^{-\rho_l}), & \hbox{\rm{if }} \rho = 1, \sum | q\sjk | = 0 \hbox{\rm{ and }} V_{12}^{l} \neq 0, \cr & \hbox{\rm{ for any $\rho_l$, }} 0 \leq \rho_l < \gamma_1 - 1, \\
O(v^{-1}), & \hbox{\rm{if }} \rho = 1, \sum | q\sjk | = 0 \hbox{\rm{ and }} V_{12}^{l} = 0.
\end{cases}
\phantom{abcdef}
\end{eqnarray}

where $\gamma_2 :=
\begin{cases}
\gamma_1, & \hbox{\rm{if }} q_{12} = 0 \hbox{\rm{ and }} V_{12}^{l} \neq 0,
\\
\gamma_D + \mu, & \hbox{\rm{if }} q_{12} \neq 0, \hbox{\rm{ and }} V_{12}^{l} \neq 0,
\\
2, & \hbox{\rm{if }} V_{12}^{l} = 0.
\end{cases}$

\end{theorem}

\begin{remark}\label{RemarkTotalPotential}{\rm
Note that the first term in the right-hand side of \eqref{Art1eq 3.24} can be written as
$$
i \int_{-\infty}^\infty d\tau \left( \left( \frac{\partial V_{12} }{\partial x_l} \right) (x + \tau \hat v) \Phi_{12}, \, \Psi_{12} \right),
$$
where $V_{12}=V_{12}^{vs}+V_{12}^s+V_{12}^l,$ and the derivative $\frac{\partial V_{12}}{\partial x_l}$ is taken in distribution sense. This shows that the high-velocity limit of $v(i[S^{D}, \bp_l] \Phi_{\bsv}, \Psi_{\bsv})$ is independent of the decomposition of the potential $V$ into the part $V^{VS}+V^S,$ that is of short range under the constant electric field $\bE$ and the part $V^L$ that is long-range; that is used for the definition of the Dollard scattering operator \eqref{Art1eq 1.29}.
}
\end{remark}

\begin{dem}

The scattering operator can be expressed as $S^{D} = (\Omega_+^{D,G,v} I^{-1}_{G,v,0,+ \infty})^*(\Omega_-^{D,G,v} I^{-1}_{G,v,0,- \infty}) = $  \\ $ I_{G,v}(\Omega_+^{D,G,v})^* \Omega_-^{D,G,v},$  by \eqref{Art1eq 1.29} and \eqref{Art1eq 3.24 -10}.
Noting that $[S^{D},\bp_l] = [S^{D},\bp_l-\mu_{12} v_l] = [S^{D}-I_{G,v},\bp_l-\mu_{12} v_l]$ and $(\bp_l- \mu_{12} v_l) \Phi_\bsv = (\bp_l \Phi_0)_\bsv$ where $\bp_l$ and $v_l$ are the l-th components of the relative momentum and the velocity $\bv$ of the chosen pair $(1,2)$, respectively. Since $\Omega_\pm^{D,G,v}$ are partially isometric and, by Duhamel formula, \eqref{Art1eq 2.1} and \eqref{Art1eq 2.3}, as in the proof of Lemma \ref{Art2le 2.3},
$$
i(S^{D}-I_{G,v}) \Phi_\bsv
= I_{G,v} i \left( \Omega_+^{D,G,v} -  \Omega_-^{D,G,v} \right)^{*}\Omega_-^{D,G,v} \Phi_\bsv
= I_{G,v} \int_{-\infty}^{+\infty} dt \ \left( U^{D,G,v}(t) \right)^{*} V_t(\tbx) e^{-iHt} \Omega_-^{D,G,v} \Phi_\bsv,
$$ with $\tbx$ defined as \eqref{Art1eq LX} and
$V_t = V_{3,t} + V_{12,t}$ where
\begin{eqnarray*}
V_{3,t}(\tbx)
&=& \sum\limits_{j < k, 3 \leq k \leq N}
\left[
V^{vs}_{jk}(\tbx\sk-\tbx\sj)
+ V_{jk}^{l} (\tbx\sk - \tbx\sj) - V^{l}_{jk}(t \, \bp\sjk/\mu\sjk - \be_1 q\sjk E t^2/(2\mu\sjk)) \right]
\\
& &
+ \sum\limits_{j < k, 3 \leq k \leq N}^E \left[ V^{s}_{jk}(\tbx\sk-\tbx\sj) - V^{s}_{jk}(\bv\sjk t + \be_1 q\sjk E t^2/(2\mu\sjk))\right]
\end{eqnarray*} and
\begin{eqnarray}
V_{12,t} &=& V^{vs}_{12}(\bx)
+ V^{l}_{12} (\bx) - V^{l}_{12}(t \, \bp/\mu_{12} - \be_1 q_{12} E t^2/(2\mu_{12})) \nonumber \\
& &
+ V^{s}_{12}(\bx) - V^{s}_{12}(\bv t + \be_1 q_{12} E t^2/(2\mu_{12})).  \label{Art2le 2.3 eq7}
\end{eqnarray}

Thus we have
\beq \label{Art1eq 2.29}
v\left(i[S^{D}, \bp_l] \Phi_{\bsv}, \Psi_{\bsv}\right)
 = I_{G,v} \left(I(v)+R(v)\right)
\ene
with
\begin{eqnarray}\label{art1eq 2.29b}
I(v)
& = &  \int_{-\infty}^{+\infty} d\tau \, l_v (\tau),
\end{eqnarray}
where
\begin{eqnarray}
l_v (vt)
& = & \left(V_{12,t}(\bx) e^{-itH\scero} \tilde{U}_{D}(t) \left(\bp_l\Phi_{0}\right)_{\bsv}, e^{-itH\scero} \tilde{U}_{D}(t) \Psi_{\bsv}\right)
\nonumber \\
&   & \ \
- \left( V_{12,t}(\bx) e^{-itH\scero} \tilde{U}_{D}(t) \Phi_{\bsv}, e^{-itH\scero} \tilde{U}_{D}(t) \left(\bp_l\Psi_{0}\right)_{\bsv}\right) \ \ \label{Art1eq 2.30}
\end{eqnarray}
and
\begin{eqnarray}
R(v)/v & = &   \int_{-\infty}^{+\infty} dt \left[ \left(V_{3,t} e^{-itH\scero} \tilde{U}_{D}(t) \left(\bp_l\Phi_{0}\right)_{\bsv}, e^{-itH\scero} \tilde{U}_{D}(t) \Psi_{\bsv}\right) \right.  \nonumber \\
&   & - \left( V_{3,t} e^{-itH\scero} \tilde{U}_{D}(t) \Phi_{\bsv}, e^{-itH\scero} \tilde{U}_{D}(t) \left(\bp_l\Psi_{0}\right)_{\bsv}\right)
+ \left( \left( e^{-iHt} \Omega_-^{D,G,v} - U^{D,G,v}(t)\right) \left(\bp_l\Phi_{0}\right)_{\bsv}, \right. \nonumber \\
&   &  \ \ \ \ V_t U^{D,G,v}(t)\Psi_{\bsv}\Big)- \left. \left(  \left( e^{-iHt} \Omega_-^{D,G,v} - U^{D,G,v}(t)\right) \Phi_{\bsv}, V_t U^{D,G,v}(t)\left(\bp_l\Psi_{0}\right)_{\bsv}\right)  \right]. \label{Art1eq 2.30 05}
\end{eqnarray} In the derivation of \eqref{Art1eq 2.30} and \eqref{Art1eq 2.30 05} we used that $\tilde{U}_{G,v} (t)$ commutes with any operator.

We are going to need to translate the Dollard-type modifier \eqref{Art1eq 1.24}
\beq
\tilde{U}_{D}(\bv, t) = e^{-i \mu_{12} \bsv \cdot \bsx}
			\prod_{j=3}^N e^{-i \mu_{j} \bsv\sj \cdot \bsx\sj}
			  \:
				\tilde{U}_{D}(t)
			  \:
			e^{i \mu_{12} \bsv \cdot \bsx}
			\prod_{j=3}^N e^{i \mu_{j} \bsv\sj \cdot \bsx\sj}. \label{Art1eq 2.33 -05}
\ene

Using equations \eqref{Art1eq 2.1}-\eqref{Art0eq 3.14 05} and substituting \eqref{Art2le 2.3 eq7}
and \eqref{Art1eq 2.33 -05}
in \eqref{Art1eq 2.30},
it follows that
\beq\label{Art1eq 2.34 abc}
I(v)  =   J_1 (v) - J_2 (v) + iJ_3 (v) + iJ_4 (v),
\ene where
\begin{eqnarray}
J_1 (v) & = & \int \left( V_{12}^{vs}(\bx+\tau \hat{\bv}+\be_1 q_{12}E\tau^2/(2v^2\mu_{12})) \, e^{-i \tau\bsp^2/(2v\mu_{12})} \tilde{U}_{D}(\bv, \tau/v) \bp_l {\Phi}_{0}, \right. \nonumber
\\
&   & \ \ \left. e^{-i \tau\bsp^2/(2v\mu_{12})} \, \tilde{U}_{D}(\bv, \tau/v) {\Psi}_{0}\right) d\tau, \label{Art1eq 2.34 a}
\\
J_2 (v) & = &  \int \left( V_{12}^{vs}(\bx+\tau \hat{\bv}+\be_1 q_{12}E\tau^2/(2v^2\mu_{12})) \, e^{-i \tau\bsp^2/(2v\mu_{12})} \, \tilde{U}_{D}(\bv, \tau/v) {\Phi}_{0}, \right. \nonumber
\\
&   & \ \ \left. e^{-i \tau\bsp^2/(2v\mu_{12})} \, \tilde{U}_{D}(\bv, \tau/v) \bp_l {\Psi}_{0} \right) d\tau, \label{Art1eq 2.34 b}
\\
J_3 (v) & = & \int \left( \left( \partial V_{12}^{E,s}/\partial x_l \right) (\bx+\tau \hat{\bv}+\be_1 q_{12}E\tau^2/(2v^2\mu_{12})) \, e^{-i \tau\bsp^2/(2v\mu_{12})} \, \tilde{U}_{D}(\bv, \tau/v) {\Phi}_{0}, \right. \nonumber \\
&   & \ \ \left. e^{-i \tau\bsp^2/(2v\mu_{12})} \, \tilde{U}_{D}(\bv, \tau/v) {\Psi}_{0} \right) d\tau. \label{Art1eq 2.34 c}
\\
J_4 (v) & = & \int \left( \left( \partial V_{12}^{l}/\partial x_l \right) (\bx+\tau \hat{\bv}-\be_1 q_{12}E\tau^2/(2v^2\mu_{12})) \, e^{-i \tau\bsp^2/(2v\mu_{12})} \, \tilde{U}_{D}(\bv, \tau/v) {\Phi}_{0}, \right. \nonumber
\\
&   & \ \ \left. e^{-i \tau\bsp^2/(2v\mu_{12})} \, \tilde{U}_{D}(\bv, \tau/v) \bp_l {\Psi}_{0} \right) d\tau. \label{Art1eq 2.34 d}
\end{eqnarray}

There exists $C>0$ that uniformly bounds the following expression, for all $j<k$:
$$
\left\Vert \Phi_{\bsv} \right\Vert +
\left\Vert \left(\bp_l\Phi_0\right)_{\bsv} \right\Vert +
\left\Vert (1+\vert \tbx\sk-\tbx\sj \vert^2)^{2} \Phi_{\bsv} \right\Vert +
\left\Vert (1+\vert \tbx\sk-\tbx\sj \vert^2)^{2}\left(\bp_l\Phi_0\right)_{\bsv} \right\Vert \leq C. $$

Then,
\begin{eqnarray*}
\frac{ |R(v)| }{v}
& \leq & C \sum\limits_{j < k, 3 \leq k \leq N} \int dt \left\Vert V^{vs}_{jk}(\tbx\sk-\tbx\sj) e^{-itH\scero} \tilde{U}_{D}(t)  \prod_{j^\prime<k^\prime} f_{j^\prime k^\prime} (\bp_{\jp\kp} - \mu\sjpkp \bv\sjpkp) (1+\vert \tbx\sk-\tbx\sj \vert^2)^{-2} \right\Vert
\\
&   & + C \sum\limits_{j < k, 3 \leq k \leq N} \int dt \left\Vert \left(V^{l}_{jk}(\tbx\sk-\tbx\sj) - V^{l}_{jk}(t \, \bp\sjk/\mu\sjk - \be_1 q\sjk E t^2/(2\mu\sjk))\right) \phantom{\prod_{j^\prime<k^\prime}} \right. \\
&   & \ \ \  \left. \times e^{-itH\scero} \tilde{U}_{D}(t)  \prod_{j^\prime<k^\prime} f_{j^\prime k^\prime} (\bp_{\jp\kp} - \mu\sjpkp \bv\sjpkp) (1+\vert \tbx\sk-\tbx\sj \vert^2)^{-2} \right\Vert
\\
&   & + C \sum\limits_{j < k, 3 \leq k \leq N}^{E} \int dt \left\Vert \left(V^{s}_{jk}(\tbx\sk-\tbx\sj) - V^{s}_{jk}(\bv\sjk t + \be_1 q\sjk E t^2/(2\mu\sjk))\right) \phantom{\prod_{j^\prime<k^\prime}} \right. \\
&   & \ \ \  \left. \times e^{-itH\scero} \tilde{U}_{D}(t)  \prod_{j^\prime<k^\prime} f_{j^\prime k^\prime} (\bp_{\jp\kp} - \mu\sjpkp \bv\sjpkp) (1+\vert \tbx\sk-\tbx\sj \vert^2)^{-2} \right\Vert
\\
&   & + C \left[ \sup_{t \in \R} \left\Vert \left( e^{-iHt} \Omega_-^{D,G,v} - U^{D,G,v}(t) \right)\left(\bp_l\Phi_0\right)_{\bsv} \right\Vert + \sup_{t \in \R} \left\Vert \left( e^{-iHt} \Omega_-^{D,G,v} - U^{D,G,v}(t)\right)\Phi_{\bsv} \right\Vert \right] \\
&   & \ \ \times \left[ \sum\limits_{j < k} \int dt \left\Vert V^{vs}_{jk}(\tbx\sk-\tbx\sj) e^{-itH\scero} \tilde{U}_{D}(t)  \prod_{j^\prime<k^\prime} f_{j^\prime k^\prime} (\bp_{\jp\kp} - \mu\sjpkp \bv\sjpkp) (1+\vert \tbx\sk-\tbx\sj \vert^2)^{-2} \right\Vert \right. \nonumber
\end{eqnarray*}
\begin{eqnarray*}
&    & \ \ \ \ + \sum\limits_{j < k} \int dt \left\Vert \left(V^{l}_{jk}(\tbx\sk-\tbx\sj) - V^{l}_{jk}(t \, \bp\sjk/\mu\sjk - \be_1 q\sjk E t^2/(2\mu\sjk))\right) \phantom{\prod_{j^\prime<k^\prime}} \right. \nonumber \\
&    & \ \ \ \ \ \ \times \left. e^{-itH\scero} \tilde{U}_{D}(t)  \prod_{j^\prime<k^\prime} f_{j^\prime k^\prime} (\bp_{\jp\kp} - \mu\sjpkp \bv\sjpkp) (1+\vert \tbx\sk-\tbx\sj \vert^2)^{-2} \right\Vert \nonumber
\\
&    & \ \ \ \ + \sum\limits_{j < k}^{E} \int dt \left\Vert \left(V^{s}_{jk}(\tbx\sk-\tbx\sj) - V^{s}_{jk}(\bv\sjk t + \be_1 q\sjk E t^2/(2\mu\sjk))\right) \phantom{\prod_{j^\prime<k^\prime}} \right. \nonumber \\
&    & \ \ \ \ \ \ \times \left. \left. e^{-itH\scero} \tilde{U}_{D}(t)  \prod_{j^\prime<k^\prime} f_{j^\prime k^\prime} (\bp_{\jp\kp} - \mu\sjpkp \bv\sjpkp) (1+\vert \tbx\sk-\tbx\sj \vert^2)^{-2} \right\Vert \right]. \nonumber
\end{eqnarray*}

Thus, by Lemmata \ref{Art2le 2.1}, \ref{Art2le 2.2}
and \ref{Art2le 2.3}, if $V^{l}\sjk = 0$ for all $1 \leq j < k \leq N:$
\begin{eqnarray}
R(v) & = &
    \begin{cases}
        o(v^{-\rho}), & \textrm{if } 0 \leq \rho < 2\alpha-1,  \\
        O(v^{-\rho}), & \textrm{if } \rho = 2\alpha-1 < 1, \sum\limits_{j < k} \vert q\sjk \vert > 0, \\
        O(v^{-1}), & \textrm{if } \rho = 1, \sum\limits_{j < k} \vert q\sjk \vert = 0, \\
    \end{cases} \label{Art1eq 2.43b}
\end{eqnarray}

Similarly, by Lemmata \ref{Art2le 2.1}, \ref{Art2le 2.2}, \ref{Art2le 3.4}
and \ref{Art2le 2.3}, if $V^{l}\sjk \neq 0$ for some $1 \leq j < k \leq N:$
\begin{eqnarray}
 R(v)
&=&
      \begin{cases}
        O \left(v^{1-2\min \{\alpha, \, \sigma\sjk  \, | \, 1 \leq j < k \leq N \}} \right), & \textrm{if } \sum\limits_{j < k} \vert q\sjk \vert > 0, \\
        O(v^{-1}), & \textrm{if } \sum\limits_{j < k} \vert q\sjk \vert = 0,
      \end{cases} \nonumber
\\
& = &
    \begin{cases}
        o(v^{-\rho}), & \textrm{if } 0 \leq \rho < 2\min \{\alpha, \, \sigma\sjk  \, | \, 1 \leq j < k \leq N \}-1,  \\
        O(v^{-\rho}), & \textrm{if } \rho = 2\min \{\alpha, \, \sigma\sjk  \, | \, 1 \leq j < k \leq N \}-1 < 1, \\
        O(v^{-1}), & \textrm{if } \rho = 1, \sum\limits_{j < k} \vert q\sjk \vert = 0.
    \end{cases} \label{Art1eq 2.43b lr}
\end{eqnarray}

Now, let us compute the following: $ \lim_{v \to \infty} v\left(i[S^{D}, \bp_l] \Phi_{\bsv}, \Psi_{\bsv}\right),$ using \eqref{Art1eq 2.29}:
\begin{eqnarray*}
\lim_{v \to \infty} I_{G,v} =
\exp \left( - i \sum\limits_{j < k}^E \lim_{v \to \infty} \int^{\infty}_{-\infty} ds\; V^{s}_{jk}(\bv\sjk s + \be_1 q\sjk E s^2/(2\mu\sjk))\right)
=1.
\end{eqnarray*}

We have used the Lebesgue dominated convergence theorem: There exist $0 < \delta_1, \delta_2 \leq 1$ such that $\vert \bv\sjk s + \be_1 q\sjk E s^2/(2\mu\sjk) \vert$ \phantom{} $\geq \sqrt{\delta_1 \vert \bv\sjk t \vert^2 + \delta_2 (q\sjk E / (2\mu\sjk))^2 t^4}$, since, $\vert \hat{\bv}\sjk \cdot \hat{\bE} \vert \leq \delta < 1$, by \eqref{Art1eq 2.13 b}, when $q_{12} = 0$, we can take $\delta_1=\delta_2=1,$ and if $q_{12} \neq 0,$ we use $\delta_1=\delta_2=1-\delta.$ We can estimate $V^{s}_{jk}(\bv\sjk s + \be_1 q\sjk E s^2/(2\mu\sjk))$ as follows:
\begin{eqnarray*}
\left\vert V^{s}_{jk}(\bv\sjk s + \be_1 q\sjk E s^2/(2\mu\sjk)) \right\vert
& \leq & C \left( 1 + \left\vert \bv\sjk s + \be_1 q\sjk E s^2/(2\mu\sjk) \right\vert\right)^{-\gamma} \\
& \leq & C \left(1+\delta_1 \bv\sjk^2 s^2 + \delta_2 |q\sjk E / (2\mu\sjk) |^2 s^4 \right)^{-\gamma/2} \\
& \leq & C (1+s^{-2\gamma}).
\end{eqnarray*}
This last term is integrable in $\R
$ because $1/2 < \gamma \leq 1$.

Note that pointwise in $\tau$,
\begin{eqnarray}\label{Art1eq 2.35}
\lim_{v \to \infty} l_v (\tau)
&=&( V_{12}^{vs}(\bx+\tau \hat{\bv}) \, (\bp_l \Phi_{12}), \, \Psi_{12}) - ( V_{12}^{vs}(\bx+\tau \hat{\bv}) \, \Phi_{12}, \, \left(\bp_l \Psi_{12}\right))
\nonumber \\
& &
+i \left( \left( \frac{\partial V_{12}^{s}}{\partial x_l} \right) (\bx+\tau \hat{\bv}) \, \Phi_{12}, \Psi_{12} \right)
+i \left( \left( \frac{\partial V_{12}^{l}}{\partial x_l} \right) (\bx+\tau \hat{\bv}) \, \Phi_{12}, \Psi_{12} \right).
\end{eqnarray}

We want to compute $\lim_{v \to \infty} I(v)$, by \eqref{art1eq 2.29b}, \eqref{Art1eq 2.35} and the Lebesgue dominated convergence theorem, thus showing the rate of convergence when $\rho=0$ in \eqref{Art1eq 2.5}:
\begin{eqnarray}\label{Art1eq 2.26a}
\lim_{v \to \infty} I(v)
&=& \int_{-\infty}^{\infty} \Big[( V_{12}^{vs}(\bx+\tau \hat{\bv}) \, (\bp_l \Phi_{12}), \, \Psi_{12}) - ( V_{12}^{vs}(\bx+\tau \hat{\bv}) \, \Phi_{12}, \, \left(\bp_l \Psi_{12}\right)) \nonumber \\
& &
\ \left. +i \left( \left( \frac{\partial V_{12}^{s}}{\partial x_l} \right) (\bx+\tau \hat{\bv}) \, \Phi_{12}, \Psi_{12} \right)
+i \left( \left( \frac{\partial V_{12}^{l}}{\partial x_l} \right) (\bx+\tau \hat{\bv}) \, \Phi_{12}, \Psi_{12} \right)
\right] d\tau,
\end{eqnarray} this means in terms of the $J_1, J_2, J_3, J_4$
functions that
\begin{eqnarray}
\lim_{v \to \infty} J_1 (v) & = & \int \left( V_{12}^{vs}(\bx+\tau \hat{\bv}) (\bp_l \Phi_{12}), \Psi_{12}\right) d\tau, \label{Art1eq 2.26a a} \\
\lim_{v \to \infty} J_2 (v) & = &  \int \left( V_{12}^{vs}(\bx+\tau \hat{\bv}) \, \Phi_{12}, \left(\bp_l \Psi_{12}\right)\right) d\tau, \label{Art1eq 2.26a b} \\
\lim_{v \to \infty} J_3 (v)  & = & \int \left( \left( \partial V_{12}^{E,s}/\partial x_l \right) (\bx+\tau \hat{\bv}) \, \Phi_{12}, \Psi_{12} \right) d\tau, \label{Art1eq 2.26a c}
\\
\lim_{v \to \infty} J_4 (v)  & = & \int \left( \left( \partial V_{12}^{l}/\partial x_l \right) (\bx+\tau \hat{\bv}) \, \Phi_{12}, \Psi_{12} \right) d\tau. \label{Art1eq 2.26a d}
\end{eqnarray}

To justify the use of the Lebesgue dominated convergence theorem observe that $\frac{\partial V_{12}^{s}}{\partial x_l}$
and $\frac{\partial V_{12}^{l}}{\partial x_l}$ are
very short-range. By \eqref{Art1eq 2.30}
and Lemma \ref{Art2le 2.1}:
\begin{eqnarray*}
\vert l_v (\tau) \vert
& \leq & C \left\Vert V^{vs}_{12}(\bx) e^{-i(\tau/v) H\scero} \tilde{U}_{D}(\tau/v)  \prod_{j^\prime<k^\prime} f_{j^\prime k^\prime} (\bp_{\jp\kp} - \mu\sjpkp \bv\sjpkp) (1+\vert \bx \vert^2)^{-2} \right\Vert \\
&  & + C \left\Vert \frac{\partial V^{s}_{12}}{\partial x_l}(\bx) e^{-i(\tau/v) H\scero} \tilde{U}_{D}(\tau/v)  \prod_{j^\prime<k^\prime} f_{j^\prime k^\prime} (\bp_{\jp\kp} - \mu\sjpkp \bv\sjpkp) (1+\vert \bx \vert^2)^{-2} \right\Vert \\
&  & + C \left\Vert \frac{\partial V^{l}_{12}}{\partial x_l}(\bx) e^{-i(\tau/v) H\scero} \tilde{U}_{D}(\tau/v)  \prod_{j^\prime<k^\prime} f_{j^\prime k^\prime} (\bp_{\jp\kp} - \mu\sjpkp \bv\sjpkp) (1+\vert \bx \vert^2)^{-2} \right\Vert \\
& \leq & C h_{12}(\vert \tau \vert),
\end{eqnarray*} where $h_{12} \in L^1((0,\infty)).$

Let us find the rate of convergence of \eqref{Art1eq 2.26a} when $\rho > 0$ in \eqref{Art1eq 2.5}. We estimate the rate of convergence of, $J_1$, the first term in the right-hand side of \eqref{Art1eq 2.34 abc} (i.e. \eqref{Art1eq 2.34 a}) to its limit.
From \eqref{Art1eq 2.34 a} and \eqref{Art1eq 2.26a a} we have:
\begin{eqnarray*}\label{Art1eq 2.37}
J_1 (v) - \lim_{v \to \infty} J_1 (v)
& = & \int_{-\infty}^{-\infty} d\tau \left(  V_{12}^{vs}(\bx+\tau \hat{\bv}+\be_1 q_{12}E\tau^2/(2v^2\mu_{12})) \, e^{-i \tau\bsp^2/(2v\mu_{12})} \, \tilde{U}_{D}(\bv, \tau/v) (\bp_l {\Phi}_{0}), \right. \nonumber \\
&   & \quad \left. e^{-i \tau\bsp^2/(2v\mu_{12})} \, \tilde{U}_{D}(\bv, \tau/v) {\Psi}_{0} \right) - \int_{-\infty}^{-\infty} d\tau \left( V_{12}^{vs}(\bx+\tau \hat{\bv}) (\bp_l \Phi_{12}), \Psi_{12}\right) \nonumber
\end{eqnarray*}
\begin{eqnarray*}
& = & \int d\tau \left[ \left( V_{12}^{vs}(\bx+\tau \hat{\bv}) e^{-i \bsp \cdot \bse_1 q_{12}E\tau^2/(2v^2\mu_{12})}\, e^{-i \tau\bsp^2/(2v\mu_{12})} \, \tilde{U}_{D}(\bv, \tau/v) (\bp_l {\Phi}_{0}), \right. \right. \nonumber \\
&   & \quad \left. \left. e^{-i \bsp \cdot \bse_1 q_{12}E\tau^2/(2v^2\mu_{12})} \, e^{-i \tau\bsp^2/(2v\mu_{12})} \, \tilde{U}_{D}(\bv, \tau/v) {\Psi}_{0} \right) \right. \nonumber \\
&   & \quad  \left. - \left( V_{12}^{vs}(\bx+\tau \hat{\bv}) e^{-i \bsp \cdot \bse_1 q_{12}E\tau^2/(2v^2\mu_{12})}\, e^{-i \tau\bsp^2/(2v\mu_{12})} \, \tilde{U}_{D}(\bv, \tau/v) (\bp_l \Phi_{0}), \Psi_{0} \right) \right] \nonumber \\
&   & + \int d\tau \left[ - \left( (\bp_l \Phi_{0}), V_{12}^{vs}(\bx+\tau \hat{\bv}) \Psi_{0}\right) \right. \nonumber \\
&   & \quad \left. + \left( e^{-i \bsp \cdot \bse_1 q_{12}E\tau^2/(2v^2\mu_{12})}\, e^{-i \tau\bsp^2/(2v\mu_{12})} \, \tilde{U}_{D}(\bv, \tau/v) (\bp_l \Phi_{0}), V_{12}^{vs}(\bx+\tau \hat{\bv}) \Psi_{0} \right) \right]. \nonumber
\end{eqnarray*}

The latter calculations suggest us to define:
\begin{eqnarray}
h^{(1)}_\bsv
& = & \left( V_{12}^{vs}(\bx+\tau \hat{\bv}) e^{-i \bsp \cdot \bse_1 q_{12}E\tau^2/(2v^2\mu_{12})}\, e^{-i \tau\bsp^2/(2v\mu_{12})} \, \tilde{U}_{D}(\bv, \tau/v) (\bp_l \Phi_{0}), \right. \nonumber \\
&   & \quad \left. \left( e^{-i \bsp \cdot \bse_1 q_{12}E\tau^2/(2v^2\mu_{12})} \, e^{-i \tau\bsp^2/(2v\mu_{12})} \, \tilde{U}_{D}(\bv, \tau/v) - I \right) \Psi_{0} \right), \label{Art1eq 2.38} \\
h^{(2)}_\bsv & = & \left( \left( e^{-i \bsp \cdot \bse_1 q_{12}E\tau^2/(2v^2\mu_{12})}\, e^{-i \tau\bsp^2/(2v\mu_{12})} \, \tilde{U}_{D}(\bv, \tau/v) - I \right) (\bp_l \Phi_{0}), V_{12}^{vs}(\bx+\tau \hat{\bv}) \Psi_{0} \right). \label{Art1eq 2.39}
\end{eqnarray} With this notation
\beq\label{Art1eq 2.39 15}
 J_1 (v) - \lim_{v \to \infty} J_1 (v) = \int d\tau \left( h^{(1)}_\bsv + h^{(2)}_\bsv \right).
\ene

Let us analyze the rate of convergence of $h^{(1)}_\bsv$.
On one hand, with $t = \tau/v:$
\begin{eqnarray*}
\left\Vert (e^{-i \bsp_1 q_{12} E \tau^2 /(2\mu_{12}v^2)} e^{-i \bsp^2 \tau / (2\mu_{12}v)}
\tilde{U}_{D}(\bv, \tau/v) - I) \Psi_{0} \right\Vert^2
&\leq& \left[C  \Big\vert \tau / v \Big\vert  \left(1 + \Big\vert \tau/v \Big\vert\right) \right]^2,
\nonumber
\end{eqnarray*}
On the other hand:
\begin{eqnarray*}
\left\Vert (e^{-i \bsp_1 q_{12} E \tau^2 / (2\mu_{12}v^2) } e^{-i \bsp^2 \tau / (2\mu_{12}v)} \, \tilde{U}_{D}(\bv, \tau/v) - I) \Psi_{0} \right\Vert
&\leq& 2 \left\Vert \Psi_{12} \right\Vert. \nonumber
\end{eqnarray*}

Consider two cases, with $0 < a \leq 1$ in mind:
\begin{itemize}
\item [$(a)$] $\vert \tau / v \vert < 1$:
Clearly we have that $\vert \tau / v \vert^a \geq  \vert \tau / v \vert$, therefore:
\begin{eqnarray}
\left\Vert (e^{-i \bsp_1 q_{12} E \tau^2 / (2\mu_{12}v^2) } e^{-i \bsp^2 \tau / (2\mu_{12}v)} \, \tilde{U}_{D}(\bv, \tau/v) - I) \Psi_{0} \right\Vert
\leq C  \Big\vert \tau/v \Big\vert \left(1 + \Big\vert \tau/v \Big\vert\right)
\leq C  \Big\vert \tau/v \Big\vert^a. \phantom{espacio} \label{Art1eq 2.38c}
\end{eqnarray}

\item [$(b)$] $\vert \tau / v \vert \geq 1$:
In this case $\vert \tau / v \vert^a \geq  1$, thus:
\begin{eqnarray}
\left\Vert (e^{-i \bsp_1 q_{12} E \tau^2 / (2\mu_{12}v^2) } e^{-i \bsp^2 \tau / (2\mu_{12}v)} \, \tilde{U}_{D}(\bv, \tau/v) - I) \Psi_{0} \right\Vert \leq C  \leq C  \Big\vert \tau/v \Big\vert^a. \label{Art1eq 2.38e}
\end{eqnarray}

\end{itemize}

Now we study $\vert h_{\bsv}^{(1)}(\tau) \vert$'s decay as $v \to \infty$ applying Lemma \ref{Art2le 2.1}, and \eqref{Art1eq 2.38c}, \eqref{Art1eq 2.38e} with $a=\rho$:
\begin{eqnarray*}
\vert h_{\bsv}^{(1)}(\tau) \vert
&\leq& C \Big\vert \tau/v \Big\vert^\rho \left\Vert V_{12}^{vs}\left(\bx + \tau \, \bp/(\mu_{12}v)+\be_1 q_{12} E\tau^2/(2v^2\mu_{12})\right) \, \tilde{U}_{D}(\tau/v) \phantom{\prod_{j^\prime<k^\prime}} \right. \nonumber \\
&    & \left. \times  \prod_{j^\prime<k^\prime} f_{j^\prime k^\prime} (\bp_{\jp\kp} - \mu\sjpkp \bv\sjpkp) (1 + \vert \bx \vert^2)^{-2}  \right\Vert.
\end{eqnarray*}
Then
\beq
v^\rho \vert h_{\bsv}^{(1)}(\tau) \vert
\leq
C \vert \tau \vert^\rho h_{12}(\vert \tau \vert) \in L^1(-\infty, \infty). \label{Art1eq 2.40b}
\ene

Hence, for $\rho=1$
$$ v \int \vert h_{\bsv}^{(1)}(\tau) \vert d\tau \leq C.$$

For $0 \leq \rho < 1,$ by Lebesgue dominated convergence theorem $$ \lim_{v \to \infty} v^\rho \int h_{\bsv}^{(1)} (\tau) d\tau = \int \lim_{v \to \infty} v^\rho h_{\bsv}^{(1)} (\tau) d\tau = 0,$$ where we used that $ \lim_{v \to \infty} v^\rho h_{\bsv}^{(1)} (\tau) = 0,$ since by \eqref{Art1eq 2.38c} and \eqref{Art1eq 2.38e} with $a=1$ we have $v^\rho \vert h_{\bsv}^{(1)} (\tau) \vert \leq C \vert \tau \vert v^{\rho-1}.$

As a result
\beq\label{Art1eq 2.41}
\int_{-\infty}^{+\infty} d\tau h_{\bsv}^{(1)}(\tau) =
\begin{cases}
o(v^{-\rho}), & \textrm{if } 0 \leq \rho < 1, \\
O(v^{-1} ), & \textrm{if } \rho = 1.
\end{cases}
\ene

At this moment, we turn our attention to the rate of convergence of $h^{(2)}_\bsv$. When $\left| \bx + \tau \hat{\bv} \right| \leq \vert \tau \vert / 2$, we have $| \bx | \geq |\tau |  - \left| \bx + \tau \hat{\bv} \right| \geq |\tau |/2.$ With the last inequality we can estimate the second factor in the scalar product of \eqref{Art1eq 2.39}. Let $g$ be a $C_0^\infty (\R^n)$ such that $g(\bp)\hat{\psi}_{12}=\hat{\psi}_{12}.$ By \eqref{Art1eq 2.38c} and \eqref{Art1eq 2.38e}:
\begin{eqnarray}\label{Art1eq 2.42}
v^\rho \int_{-\infty}^{\infty} d\tau \, \vert h_{\bsv}^{(2)}(\tau) \vert
&\leq& C \int_{-\infty}^{+\infty} d \tau \vert \tau \vert^{\rho}  \left( \left\Vert V_{12}^{vs}(\bx+\hat{\bv}\tau) g(\bp) F(\vert \bx+\hat{\bv}\tau \vert \geq \vert \tau \vert/2) \right\Vert \right. \nonumber \\
& & \left. + \left\Vert V_{12}^{vs}(\bx+\hat{\bv}\tau) g(\bp) \right\Vert \left\Vert F(\vert \bx\vert \geq \vert \tau \vert/2) \Psi_{12} \right\Vert \right).
\end{eqnarray}

Due to the short-range condition \eqref{Art1eq 2.5}, the first integral in \eqref{Art1eq 2.42} is finite; the fast decay in configuration space of $\Psi_{12}$ makes the second integral in \eqref{Art1eq 2.42} be bounded:
\begin{eqnarray*}
\int_{-\infty}^{\infty} d\tau \, \vert \tau \vert^\rho \left\Vert F(\vert \bx\vert \geq \vert \tau \vert/2) \Psi_{12} \right\Vert
& = & \int_{-\infty}^{\infty} d\tau \, \vert \tau \vert^\rho (1+ \vert \tau \vert)^{-3} \left\Vert (1+ \vert \tau \vert)^{3} F(\vert \bx\vert \geq \frac{\vert \tau \vert}{2}) \Psi_{12} \right\Vert \nonumber
<  \infty.
\end{eqnarray*}

Hence, for $\rho=1$
$$ v \int \vert h_{\bsv}^{(2)}(\tau) \vert d\tau \leq C,$$ and for $0 \leq \rho < 1,$ by Lebesgue dominated convergence theorem $$ \lim_{v \to \infty} v^\rho \int h_{\bsv}^{(2)} (\tau) d\tau = \int \lim_{v \to \infty} v^\rho h_{\bsv}^{(2)} (\tau) d\tau = 0,$$ where we used that $ \lim_{v \to \infty} v^\rho h_{\bsv}^{(2)} (\tau) = 0,$ since by \eqref{Art1eq 2.38c} and \eqref{Art1eq 2.38e} with $a=1$ we have $ v^\rho \vert h_{\bsv}^{(2)} (\tau) \vert \leq C \vert \tau \vert v^{\rho-1}.$

As a result
\beq\label{Art1eq 2.43}
\int_{-\infty}^{+\infty} d\tau h_{\bsv}^{(2)}(\tau) =
\begin{cases}
o(v^{-\rho}), & \textrm{if } 0 \leq \rho < 1, \\
O(v^{-1} ), & \textrm{if } \rho = 1.
\end{cases}
\ene

We have just estimated the rate of convergence of $J_1$, the first term in the right-hand side of \eqref{Art1eq 2.34 abc}. Since $\hat{\phi}_{12} \in C_0^\infty(B_{\mu_{12} \eta }) $, we have that $\left(p_l \hat{\phi}_{12}\right) \in C_0^\infty(B_{\mu_{12} \eta })$, therefore, we can apply the same treatment to $J_2$, the second term in the right-hand side of \eqref{Art1eq 2.34 abc}. For $J_3,$
in the right-hand side of \eqref{Art1eq 2.34 abc},
when we estimate the term with $\partial / \partial x_l V_{12}^{E,s}$ we have that $$ (1+\vert x \vert)^{\rho_s} \vert \partial / \partial x_l V_{12}^{E,s} (x) \vert \leq C (1+\vert x \vert)^{-1-\alpha+\rho_s}$$ satisfies the very short-range condition if $\rho_s < \alpha \leq 1.$ Nevertheless, when $q_{12} \neq 0$, we do not have an extra error term of the form $o(v^{-\rho_s})$ because in \eqref{Art1eq 2.43b} and \eqref{Art1eq 2.43b lr} $\rho < \alpha,$ for that reason one can always choose $\rho_s$ such that $\rho < \rho_s < \alpha \leq 1.$
Regarding $J_4$ in \eqref{Art1eq 2.34 abc}, we estimate the term with $\partial / \partial x_l V_{12}^{l},$ one sees that $$ (1+\vert x \vert)^{\rho_l} \vert \partial / \partial x_l V_{12}^{l} (x) \vert \leq C
\begin{cases}
(1+\vert x \vert)^{-\gamma_D-\mu+\rho_l}, & \textrm{if } q_{12} \neq 0, \\
(1+\vert x \vert)^{-\gamma_1+\rho_l}, & \textrm{if } q_{12} = 0,
\end{cases}
$$ satisfies the very short-range condition if $\rho_l <
\begin{cases}
\gamma_D+\mu-1, & \textrm{if } q_{12} \neq 0, \\
\gamma_1 - 1, & \textrm{if } q_{12} = 0.
\end{cases}
$ Therefore, we have another error term of the form \beq\label{Art1eq 2.43 10} o(v^{-\rho_l}).\ene

Moreover, when there is at least one pair with non-zero relative charge, we have to estimate the following error, see \eqref{Art1eq 3.20 05} and \eqref{Art1eq 2.29}.
In this case, $\rho < 1,$ and $-(2\gamma-1) \leq -(2\alpha-1) \leq -\rho,$
where $\gamma$ is as in Definition \ref{Art2df 1.2}.
By equation \eqref{Art1eq 2.13 b}:
\begin{eqnarray}
\left\vert I_{G,v} - 1 \right\vert
& \leq & \sum\limits_{j < k}^E
\int^\infty_{-\infty} ds\; \vert V^{s}_{jk}(\bv\sjk s + \be_1 q\sjk E s^2/(2\mu\sjk)) \vert \nonumber
\leq
C \begin{cases}
            v^{-(2\gamma-1)}, & \textrm{if } 1/2 < \gamma < 1, \\
            \frac{\ln v}{v}, & \textrm{if } \gamma = 1,
        \end{cases}
\\
&=&
    \begin{cases}
        o(v^{-\rho}), & \textrm{if } 0 \leq \rho < 2\min \{\alpha, \, \sigma\sjk  \, | \, 1 \leq j < k \leq N \}-1,  \\
        O(v^{-\rho}), & \textrm{if } \rho = 2\min \{\alpha, \, \sigma\sjk  \, | \, 1 \leq j < k \leq N \}-1 < 1.
    \end{cases}
\label{Art1eq 2.43 15}
\end{eqnarray}

Finally, to prove the convergence rate given in \eqref{Art1eq 3.24} we sum the terms, corresponding to $I(v)$, $R(v)$, $I_{G,v}$, respectively, in \eqref{Art1eq 2.29}, recalling \eqref{Art1eq 2.43b}, \eqref{Art1eq 2.43b lr}, \eqref{Art1eq 2.39 15}, \eqref{Art1eq 2.41}, \eqref{Art1eq 2.43}, \eqref{Art1eq 2.43 15}
and taking in consideration
\eqref{Art1eq 2.43 10}
with the highest possible values of
$\rho_l$ in order to have the optimal error rate in all the cases enounced in Theorem \ref{Art2tr 2.1}.
\end{dem}

The following reconstruction formula is of independent interest.
\begin{theorem}\label{Art1tr 3.1}
Assume the same hypothesis as in Theorem \ref{Art2tr 2.1},
Then
\begin{eqnarray}\label{Art1eq 3.23}
& & v(i[S^{D}-I_{G,v}] \Phi_{\bsv}, \Psi_{\bsv})
- I_{G,v} \int_{-\infty}^{\infty} v dt \left( \left( V_{12}^{s} (\bx) - V_{12}^{s} (\bv t + \be_1 q_{12} E t^2/(2\mu_{12}))
+ V^{l}_{12} (\bx) \right. \right. \nonumber \\
& &
\left. \left. - V^{l}_{12}(t \, \bp/\mu_{12} - \be_1 q_{12} E t^2/(2\mu_{12}))
\right)
e^{-itH\scero} \tilde{U}_{D}(t) \Phi_{\bsv}, e^{-itH\scero} \tilde{U}_{D}(t) \Psi_{\bsv}
\right)
= \int_{-\infty}^{\infty} d\tau (V_{12}^{E,\,vs}(\bx + \tau \hat{\bv}) \Phi_{12}, \Psi_{12})
\nonumber \\
&  &
+
\begin{cases}
    o(v^{-\rho}), & \hbox{\rm{if }} 0 \leq \rho < 2\min \{\alpha, \, \sigma\sjk  \, | \, 1 \leq j < k \leq N \}-1,  \\
    O(v^{-\rho}), & \hbox{\rm{if }} \rho = 2\min \{\alpha, \, \sigma\sjk  \, | \, 1 \leq j < k \leq N \}-1 < 1, \\
    O(v^{-1}), & \hbox{\rm{if }} \rho = 1, \hbox{\rm{ and }} \sum | q\sjk | = 0.
\end{cases}
\end{eqnarray}
\end{theorem}

\begin{dem}
The left hand side of \eqref{Art1eq 3.23} can be written as equal to the right hand side of \eqref{Art1eq 2.29} exactly with the same $I_{G,v}$ but with $$I(v) = v \int_{-\infty}^{+\infty} dt  \left(V^{vs}_{12}(\bx) e^{-itH\scero} \tilde{U}_{D}(t) \Phi_{\bsv}, e^{-itH\scero} \tilde{U}_{D}(t) \Psi_{\bsv}\right),$$ and, with the same $V_{3,t}$ and $V_{t}$ as in the proof of Theorem \ref{Art2tr 2.1},
\begin{eqnarray*}
R(v)/v & = &   \int_{-\infty}^{+\infty} dt \left[ \left(V_{3,t} e^{-itH\scero} \tilde{U}_{D}(t) \Phi_{\bsv}, e^{-itH\scero} \tilde{U}_{D}(t) \Psi_{\bsv}\right) \right.  \nonumber \\
&   &  \ \ \ \ + \left. \left(  \left( e^{-iHt} \Omega_-^{D,G,v} - U^{D,G,v}(t)\right) \Phi_{\bsv}, V_t U^{D,G,v}(t)\Psi_{\bsv}\right)  \right].
\end{eqnarray*}
The convergence rate of $I(v)$ is computed like that one of $J_1$, see equations \eqref{Art1eq 2.34 a}, \eqref{Art1eq 2.38}, \eqref{Art1eq 2.39}, \eqref{Art1eq 2.41}, \eqref{Art1eq 2.43}. $R(v)$ and $I_{G,v}$ are estimated like in \eqref{Art1eq 2.43b lr} and \eqref{Art1eq 2.43 15}, respectively.
\end{dem}

\hbox{\bf Proof of Theorem \ref{Art1te 1.2}:}

Let us consider the states $\Phi \sim \hat{\phi}_{12}(p) \hat{\phi}_3(p_3, \ldots, p_N),$ $\Psi \sim \hat{\psi}_{12}(p) \hat{\phi}_3(p_3, \ldots, p_N),$ such that $\hat{\phi}_{12}, \hat{\psi}_{12} \in C_0^\infty (\R^n)$ and $\hat\phi_3$ is like in \eqref{Art1eq 3.2}. Let $\by$ be an element of a two dimensional subspace of $\Rn,$ for instance, we associate each $\by=(y_1,y_2) \in \R^2$ with the vector $y\suno \be\suno + y\sdos \be\sdos \in \R^n.$ We express by
\beq\label{Art1eq 2.45}
\Phi^\bsy=e^{-i \bsp\cdot\bsy} \Phi \Leftrightarrow \phi^\bsy = \phi_{12}(\bx-\by) \phi_3(\bx_3, \ldots, \bx_N), \quad \Psi^\bsy=e^{-i \bsp\cdot\bsy} \Psi \Leftrightarrow \psi^\bsy = \psi_{12}(\bx-\by)\phi_3(\bx_3, \ldots, \bx_N), 
\ene
the states, translated in the configuration space by $\by$, considered as an vector in $\R^n$.

Suppose that $V^i = V^{VS, \, i}+V^{S, \, i}+V^{L, \, i} \in \Vscr_{VSR}+\Vscr_{SR}+\Vscr_{LR}, \, i=1,2,$ and that $S^D(V^{L, \, 1}; V^{VS, \, 1} + V^{S, \, 1}) = S^D(V^{L, \, 2}; V^{VS, \, 2} + V^{S, \, 2}).$ Then, we can write the potentials $V^i, \, i=1,2,$
$$
V^i
=
\sum\limits_{1 \leq j < k \leq N} V\sjk^{i} (\tbx\sk - \tbx\sj),
\quad \quad
V\sjk^{i}
=
V\sjk^{vs, \, i} + V\sjk^{s, \, i} + V\sjk^{l, \, i},
$$
with, for all $1 \leq j < k \leq N,$ $V\sjk^{vs, \, i} \in \Vscr_{VSR},$ $V\sjk^{s, \, i} \in \Vscr_{SR},$ and $V\sjk^{l, \, i} \in \Vscr_{LR}.$

It is enough to prove uniqueness for the pair $(1,2).$ Let us assume $q_{12} \neq 0,$ the other case is similar and simpler. Note that as $q_{12} \neq 0,$ $V_{12}^{vs, \, i} \in \Vscr_{E, \, vs},$ $V_{12}^{s, \, i} \in \Vscr_{E, \, s},$ and $V_{12}^{l, \, i} \in \Vscr_{E, \, l}.$ We define
\beq\label{Art1eq 2.44 b}
\begin{cases}
Q_{12}^{vs}(\bx) &= V_{12}^{vs,2}(\bx) - V_{12}^{vs,1}(\bx), \\
Q_{12}^{s}(\bx) &= V_{12}^{s,2}(\bx) - V_{12}^{s,1}(\bx), \\
Q_{12}^{l}(\bx) &= V_{12}^{l,2}(\bx) - V_{12}^{l,1}(\bx), \\
Q_{12}(\bx) &= Q_{12}^{vs}(\bx) + Q_{12}^{s}(\bx) + Q_{12}^{l}(\bx).
\end{cases}
\ene

With $\Phi^\bsy$ and $\Psi^\bsy$ as in \eqref{Art1eq 2.45}, and $\bp\suno = \bp \cdot \be\suno,$ the function $f:\R^2 \to \mathbb{C}$ is defined as
\beq\label{Art1eq 2.46 b}
f(\by) := f_1(\by) + f_2(\by) + f_3(\by),
\ene
where
\begin{eqnarray*}
f_1(\by)&:=& (Q_{12}^{vs} (\bx) \bp\suno \Phi^{\bsy}, \Psi^{\bsy}), \nonumber \\
f_2(\by)&:=&- (Q_{12}^{vs} (\bx) \Phi^{\bsy}, \bp\suno \Psi^{\bsy}), \nonumber \\
f_3(\by)&:=& i \left( \left( \frac{\partial Q_{12}^{s}}{\partial x\suno} + \frac{\partial Q_{12}^{l}}{\partial x\suno}\right) (\bx) \Phi^{\bsy}, \Psi^{\bsy} \right).
\end{eqnarray*}

Let us focus on $f_1$. Let $g_1$ be a $C_0^\infty (\R^n)$ function such that $g_1(p)\hat{\phi}_{12}(p) = \hat{\phi}_{12}(p),$
\begin{eqnarray}
\vert f_1(\by) \vert
&\leq& C \Vert Q_{12}^{vs} (\bx) g_1(\bp) \Vert
\label{Art1eq 2.46 1a} \\
\vert f_1(\by) \vert
&\leq& C \left( \left\Vert Q_{12}^{vs} (\bx) g_1 (\bp) F(|\bx| \geq |\by|/2) \right\Vert \right. \nonumber \\
&    & + \left. \left\Vert Q_{12}^{vs} (\bx) \bp\suno g_1 (\bp) \right\Vert \left\Vert F(|\bx| < |\by|/2 ) \phi_{12} (\bx-\by) \right\Vert \right) . \label{Art1eq 2.46 1ab}
\end{eqnarray}

Inequality \eqref{Art1eq 2.46 1a} shows that $f_1$ is bounded. By the very short range condition \eqref{Art1eq 2.5}: $$\left\Vert Q_{12}^{vs} (\bx) g_1 (\bp) F(|\bx| \geq |\by|/2) \right\Vert \in L^2(\R^2).$$
Additionally
\begin{eqnarray*}
\left\Vert F(|\bx| < |\by|/2 ) \phi_{12} (\bx-\by) \right\Vert
&=& \left\Vert \frac{1}{1+ \vert \bx-\by \vert^2} F(|\bx| < |\by|/2 ) (1+ \vert \bx-\by \vert^2)\phi_{12} (\bx-\by) \right\Vert \\
&\leq& \frac{C}{(1+ \vert \by \vert/2)^2} \in L^2(\R^2).
\end{eqnarray*}

Then, $f_1(\by) \in L^2(\R^2).$ Moreover, $f_1(\by)$ is continuous because the operator $e^{-i \bsp \cdot \bsy}$ is strongly continuous on $L^2(\R^2).$

Working with $f_2$ and $f_3$ is analogous to the case of $f_1$, remarking that \eqref{Art2eq 1.6} and \eqref{Art2eq 1.6 05}
imply that $\frac{\partial Q_{12}^{s}}{\partial x\suno} + \frac{\partial Q_{12}^{l}}{\partial x\suno} $ belongs to our very short-range class $\Vscr_{E,\,vs}$. Thus $f(\by) \in L^2(\R^2)$ and it is a bounded continuous function.

The Radon transform of $f(\by)$, for any $\bv$ in the $\by$-plane satisfying $\vert \hat{\bv} \cdot \hat{\bE} \vert < 1$, is given by
\begin{eqnarray}\label{Art1eq 2.47 b}
\tilde{f}(\hat{\bv};\by)
&:=& \int_{-\infty}^{\infty} f(\by+\tau\hat{\bv}) d\tau
= \int_{-\infty}^{\infty} \left[ (Q_{12}^{vs}(\bx+\tau\hat{\bv}) \bp\suno \Phi^{\bsy}, \Psi^{\bsy}) \right. \nonumber \\
& &
- (Q_{12}^{vs}(\bx+\tau\hat{\bv}) \Phi^{\bsy} , \bp\suno \Psi^{\bsy})
\nonumber \\
& &
\left. + i\left( \left( \frac{\partial Q_{12}^{s}}{\partial x\suno}
+ \frac{\partial Q_{12}^{l}}{\partial x\suno}
\right) (\bx+\tau\hat{\bv}) \Phi^{\bsy} , \Psi^{\bsy}\right) \right].
\end{eqnarray}

By Theorem \ref{Art2tr 2.1} applied to the pair $(1,2)$, we have that
\begin{eqnarray*}
\tilde{f}(\hat{\bv};\by)
&=& \lim_{v \to \infty} \left[ v(i[S^{D}(V^{L,\, 1};V^{VS,\,1} + V^{S,\,1}), \bp\suno] \Phi_{\bsv}^{\bsy}, \Psi_{\bsv}^{\bsy}) \right. \\
& & \left. \ \ - v(i[S^{D}(V^{L,\,2};V^{VS,\,2} + V^{S,\,2}), \bp\suno] \Phi_{\bsv}^{\bsy}, \Psi_{\bsv}^{\bsy}) \right] \\
&\equiv& 0.
\end{eqnarray*}

Then, the Plancherel formula associated with the Radon transform \cite{Helgason1999} implies that $f(\by) = 0.$ From \eqref{Art1eq 2.46 b} we have that
$$
\frac{\partial}{\partial y_1} (Q_{12} \Phi^{\bsy}, \Psi^{\bsy}) = -i f(\by).
$$
This implies that $(Q_{12} \Phi^{\bsy}, \Psi^{\bsy})$ does not depend on $y_1.$ Moreover, $\lim_{|y\suno| \to \infty} (Q_{12} \Phi^{\bsy}, \Psi^{\bsy})=0$ by \eqref{Art1eq 1.3}, \eqref{Art2eq 1.5} and \eqref{Art2eq 1.6 05}. Therefore, $(Q_{12} \Phi^{\bsy}, \Psi^{\bsy}) \equiv 0.$
In particular $(Q_{12} \Phi^{0}, \Psi^{0}) = (Q_{12} \phi_{12}, \psi_{12}) = 0,$ what implies by the density of the states $\phi_{12}, \psi_{12}$ that
$Q_{12}(\bx) \equiv 0$ a.e. We conclude that the total potential $V$ is uniquely determined by the high-velocity limit of the commutator of any Dollard scattering operator $S^{D}$ and some component of the momentum.

We consider the reconstruction problem of the total potential $V$ as in \eqref{Art1eq 1.21}, by means of Theorem \ref{Art2tr 2.1}. We assume $q_{12} \neq 0$ because the case $q_{12} = 0$ is easier. Let us compute $V_{12} := V_{12}^{vs} + V_{12}^{s} + V_{12}^{l} \in \Vscr_{VSR}+\Vscr_{SR}+\Vscr_{LR}$ from the high-velocity limit of $[S^{D},\bp\suno].$
We substitute $Q_{12}^{vs}$ by $V_{12}^{vs},$ $Q_{12}^{s}$ by $V_{12}^{s}$ and $Q_{12}^{l}$ by $V_{12}^{l}$ in \eqref{Art1eq 2.46 b}. We know $\lim_{v \to \infty} v(i[S^{D}, \bp\suno] \Phi_{\bsv}^\bsy, \Psi_{\bsv}^\bsy)$ for all $\Phi^\bsy$ and $\Psi^\bsy$ as in \eqref{Art1eq 2.45}. Then, by Theorem \ref{Art2tr 2.1} and \eqref{Art1eq 2.47 b} we reconstruct $\tilde{f}(\hat{\bv};\by)$ and by the inversion of the Radon transform \cite{Helgason1999}, we uniquely reconstruct $f(\by).$  From \eqref{Art1eq 1.3}, \eqref{Art2eq 1.5}, \eqref{Art2eq 1.6} and \eqref{Art2eq 1.6 05} $f$ is integrable along any line and $\lim_{\bsy \to \infty} ((V_{12}) \Phi^\bsy, \Psi^\bsy) = 0.$ Then we have
$$
(V_{12} \phi_{12}, \psi_{12}) = i \int_0^\infty f(y_1, 0) dy_1,
$$
in a dense set in $L^2.$ Hence $V_{12}$ is obtained almost everywhere as a function. Repeating this process for all pairs we reconstruct $V.$
\hfill \bull

\begin{remark}\label{RemarkArt1tr 3.1}{\rm
As we have already mentioned in Remark \ref{RemarkInjectivity3} the reconstruction formula \eqref{Art1eq 3.23} from Theorem \ref{Art1tr 3.1} is simpler than the formula \eqref{Art1eq 3.24} in Theorem \ref{Art2tr 2.1}. Let us show how \eqref{Art1eq 3.23} can be used. Let us suppose that $q_{12} \neq 0.$ The case $q_{12}=0$ follows in the same way. The potentials $V_{12}^{E,\,vs} \in \Vscr_{E,\,vs},$ $V_{12}^{E,\,s} \in \Vscr_{E,\,s}$, $V_{12}^{E,\,l} \in \Vscr_{E,\,l}$ are the very short-, short- and long-range potentials, respectively, for the pair $(1,2).$ Let us assume that we want to recover $V_{12}^{E,\,vs}$ knowing $V_{12}^{E,\,s},$ $V_{12}^{E,\,l}$ and the high-velocity limit of $S^{D}$ for each $\Phi^\bsy$ and $\Psi^\bsy$ as in \eqref{Art1eq 2.45}. Defining
\beq\label{Art1eq 2.46}
h(\by) = (V_{12}^{E,vs} (\bx) \Phi^\bsy, \Psi^\bsy),
\ene
using Theorem \ref{Art1tr 3.1} and inverting the Radon transform we obtain $h(\by).$ Then, we can compute $(V_{12}^{E,\,vs} \phi_{12}, \psi_{12})$ = $h(0)$ in a dense set in $L^2.$ This implies that we recover $V_{12}^{E,\,vs}$ almost everywhere as a function.
}
\end{remark}

\hbox{{\bf Acknowledgment}}
We thank Professor Patrick Joly for his kind hospitality at the project POEMS at Institut National de Recherche en Informatique et en Automatique Paris-Rocquencourt, where this work was partially done. We also greatly appreciate the comments made by the referee.

{}

\end{document}